\documentclass[submit]{eps_arxiv}

\usepackage{graphicx}
\usepackage{times}
    \usepackage{threeparttable}
\usepackage[usenames]{color}

\volumenumber{58}
\publishyear{2006}
\frompage{\pageref{firstpage}}
\topage{\pageref{finalpage}}

\shorttitle{M. Siudek \lowercase{\textit{et al.}}: Infrared Composition of The Large Magellanic Cloud}

\title{Infrared Composition of The Large Magellanic Cloud}
\author{M. Siudek$^1$, A. Pollo$^{1,2,3}$, T. T. Takeuchi$^{4}$, Y. Ita$^{5}$, D. Kato$^{6}$\thanks{Present address: Center for Low Carbon Society Strategy, Japan Science and Technology Agency, 7, Goban-cho, Chiyoda-ku, Tokyo 102-0076, Japan.}, and T. Onaka$^{7}$}
\affiliation{$^1$Center for Theoretical Physics of the Polish Academy of Sciences, Al. Lotnikow 32/46, 02-668 Warsaw, Poland\\
             $^2$Astronomical Observatory of the Jagiellonian University, Orla 171, 30-001 Cracow, Poland\\
             $^3T$National Centre for Nuclear Research, Hoza 69, 00-681 Warsaw, Poland\\
             $^4$Division of Particle and Astrophysical Science, Nagoya University, Furo-cho, Chikusa-ku, Nagoya 464-8602, Japan\\
             $^5$Astronomical Institute, Tohoku University, 6-3 Aramaki Aoba, Aoba-ku, Sendai, Miyagi 980-8578, Japan\\
             $^6$Institute of Space and Astronautical Science, Japan Aerospace Exploration Agency, 3-1-1, 
				Yoshino-dai, Chuo-ku, Sagamihara, Kanagawa 252-5210, Japan\\             
             $^7$Department of Astronomy, The University of Tokyo, Bunkyo-ku, Tokyo 113-0033, Japan
             }
\received{November 17, 2011}\revised{September 2, 2012}\accepted{September 3, 2012}

\abstract{
The evolution of galaxies and the history of star formation in the Universe are among the most important topics in today's astrophysics. Especially, the role of small, irregular 
galaxies in the star-formation history of the Universe is not yet clear. 
Using the data from the AKARI IRC survey of the Large Magellanic Cloud at 3.2, 7, 11, 15, and 24 $\mu$m wavelengths, i.e., at the mid- and near- infrared,
we have constructed a multiwavelength catalog 
containing data from a cross-correlation with a number of other databases at different wavelengths. We present the separation of different classes of stars in the LMC 
in color-color, and color-magnitude, diagrams, and analyze their contribution to the total LMC flux, related to point sources at different infrared wavelengths.}


\begin{document}
\label{firstpage}
\maketitle
\copyrighttext{}

\section{Introduction}

Understanding the birth and evolution of galaxies is one of the most important problems of present-day extragalactic
astrophysics. 
In order to study the early stages of galaxy evolution, it is crucial to understand the process of star formation in irregular galaxies with low
metallicity, such as the Large Magellanic Cloud (LMC).


The irregular galaxies (Irr) are defined by their lack of organized optical structure [de Vaucouleurs and Freeman, 1972] and
are the most common type of galaxy in the Universe [Hunter, 2002]. In comparison with spiral galaxies they are, in general, 
smaller, less massive, less luminous and bluer [Gallagher and Hunter, 1984; Hunter and Gallagher, 1986]. Irregular galaxies are simpler and 
are less dusty and lower in metallicity and shear in their interstellar mediums [Hunter and Elmegreen, 2004, Hunter, 2002] than spirals. In the context of 
chemical evolution, these galaxies
are less evolved than spirals, as indicated by the high gas content and low abundances of heavy elements. That means that
their interstellar medium was not in such a rate processed through stars [Hunter, 1997]. Most Irr-type galaxies have on-going star formation and some are
forming stars at rates normalized to their size that are comparable, or even higher, than those for spiral galaxies [Hunter, 1997, Gallagher and Hunter, 1984]. 

Irregular galaxies are an ideal laboratory
to study various astrophysical processes influencing the evolution of large galaxies because their reaction to all internal or
environmental perturbations, such as outbursts of supernovae or interactions with other galaxies, is stronger and occurs in
shorter timescales due to their small dimensions and masses. Moreover, the absence of spiral density waves and
 a strikingly-different environment from that in spirals makes them useful for studying the star-formation process. Also, we believe that the first 
galaxies in the Universe were, in many aspects, similar to present-day irregular galaxies [Longair, 2008].
Hence, the necessity to understand the evolution and star-formation histories of this type of galaxy.


The Large Magellanic Cloud (LMC) is an ideal laboratory to study molecular clouds, interstellar matter, gas
and dust, and individual regions of star formation in the early stages of galaxy evolution, thanks to its favorable viewing angle ($35^{\circ}$)
[van der Marel and Cioni, 2001] and proximity ($\approx$50 kpc) [Alves, 2004]. Although LMC is sometimes classified as a one-armed
spiral galaxy (according to de Vaucouleurs classification it is a very late type of a spiral galaxy, SB(s)m., i. e.
an irregular spiral without a bulge) [Wilcots, 2009], the process of star formation is not limited to this particular structure, and we can
treat it as a model for the star-formation process of irregular galaxies. 
A metal-poor environment (Z $\approx$ 0.3 $Z_{\odot}$) [Luck et al.,
1998], and a low dust-to-gas ratio [Koornneef, 1982] are responsible for its ultraviolet radiation field and chemical conditions of
circumstellar materials [Shimonishi et al., 2010]. These characteristics may play a key role in the star-formation process of
populous star clusters that have no galactic counterpart [Fukui, 2005]. Detailed studies of star formation in the LMC will enable
an understanding of these processes. 

In galaxies with an active star-formation process, the dominant source of dust are stars in the final stage of evolution [Asano et al., 2011].
The asymptotic giant branch (AGB) is an evolutionary phase of low- and intermediate- mass
stars (0.8 $M_{\odot} < M < 8 M_{\odot}$) which actually expel gas and associated dust in the interstellar medium (ISM) via planetary nebula by superwind
phenomena. Due to heavy mass loss and nucleosynthesis, AGB stars tend to be the dominant source of dust in the ISM [Groenewegen et al., 2007]. 

Post-AGB stars are low- and intermediate- mass stars which rapidly evolve from the AGB towards the planetary nebula phase (this class includes planetary nebulae, PNe). 
Nowadays, there are a lot of unresolved questions about their evolution and physical processes, while nobody has
constructed a complete theoretical model fitting their global properties. Within a group there is observed a large range of completely
different photosphere abundance patterns, circumstellar geometries and kinematics. Moreover, their evolutionary phase is
relatively short and the group of known post-AGBs is rather small. Post-AGB stars are characterized by a wide range of
the electromagnetic spectrum. The central star is responsible for UV and optical emission, while the cool circumstellar
envelope emits in the infrared [van Aarle et al., 2009].

Studies of the LMC provide knowledge about young stellar objects (YSOs) which are a typical containment of stellar associations
(star-forming stellar systems). There are strong correlations between young associations with one or more H\sc{ii} \normalfont
regions
cataloged by Henize (1956) and Davies et al. (1976). Besides young bright stars there are a large amount of post-main
sequence (PMS) stars [Nota et al., 2006; Gouliermis et al., 2003], whose distribution is correlated with molecular clouds.
In the LMC, young associations younger than one million years are associated with the Giant Molecular Cloud (GMC)
[Yamaguchi et al. 2001; Fukui et al., 2008]. A correlation of ground-based, and Spitzer, observations of YSO candidates
show that, in these regions, the star-formation process is active, and also a second generation of star formation occurs.
%

During the last decade, observing capability has increased, and there have been a lot of 
survey projects of the LMC at various wavelengths. Most studies have been made at ultraviolet (UV) and visible wavelengths, but observations in
the infrared are crucial for investigating the formation and evolution of galaxies, because a significant fraction of the total
emission falls in the infrared. 
Compared with optical-UV bands, the infrared is free of problems with
dust absorption. Sensitive near-infrared surveys can derive information about the relative timescales of galaxy formation
and point what controls the star-formation rate in galaxies. Ground-based near-infrared (NIR) surveys derive about fifteen
million sources in the LMC [Kato et al., 2007] but mid-infrared (MIR) surveys detect only a few thousand sources [Egan et
al., 2003], which is too shallow to compare with observations in other bands. Also the poorer angular resolutions of MIR surveys
hamper a secure correlation with other catalogs. For the first time, the observations made by the Spitzer Space Telescope
(SST) [Werner et al., 2004], and AKARI, have a high spatial resolution. Moreover, the 11 and 15 $\mu$m wavelength bands
are unique to AKARI.

Observations provided by the Spitzer Space Telescope Legacy Survey (``Surveying the Agents of a Galaxy's Evolution'', SAGE) [Meixner, et al., 2006] have enabled 
studies of various astrophysical properties of evolved stars [Blum et al., 2006; Hora et al., 2008; Srinivasan et al., 2009], young 
stellar objects [Whitney et al., 2008], variables [Vijh et al., 2009], and massive stars [Bonanos et al., 2009]. A SAGE catalog of massive stars with accurate spectral 
types and multiwavelengths in the Large Magellanic Cloud includes
1750 objects [Bonanos et al., 2009]. Analysis of color-magnitude diagrams shows that supergiant B[e], red supergiant, and luminous blue variable (LBV), stars are the 
dominant source of the LMC radiation at all infrared wavelengths (3.6, 4.5, 5.8, 8.0, 24.0 $\mu$m).  
SAGE-Spec Legacy Program obtained 52-93 $\mu$m spectra of 48 compact far-IR sources in the LMC, which were used for classification purposes [van Loon et al., 2010]. 
Analysis of the fine-structure lines of oxygen, [O \sc{i} \normalfont]  at 63 $\mu$m and [O \sc{iii} \normalfont]  at 88 $\mu$m, enabled YSOs to be distinguished from 
evolved objects, and revealed new features about some objects in the LMC [van Loon et al., 2010]. 

Within the program HERshel Inventory of The Agents of Galaxy Evolution (HERITAGE) observations of Magellanic Clouds performed by the Herschel Space Observatory's PACS and 
SPIRE cameras in 5 bands, 100, 160, 250, 350 and 500 $\mu$m, will be used to 
investigate the life cycle of matter [Meixner et al. 2010]. Far-infrared and submillimeter emission from dust grains have enabled the effective 
investigation of the dust from agents of 
galaxy evolution, which are the ISM, the most deeply embedded YSOs, evolved massive stars, and supernovae [Meixner et al., 2010]. 
Maixner et al. [2010] constructed two dust models which fit the spectral energy distribution and reveal 
that 500 $\mu$m is in excess by 6$\%$ to 17$\%$. The presented dust model, which uses amorphous carbon and silicate optical properties for 
Galactic dust, yields realistic gas mass ratios (GDR), consistent with prior observations. The model using standard graphite and silicate optical properties for Galactic dust 
is unrealistic for the LMC dust [Meixner et al., 2010]. 

\begin{figure*}[t]
\centerline{\includegraphics[angle=0, width=0.95\textwidth,clip]{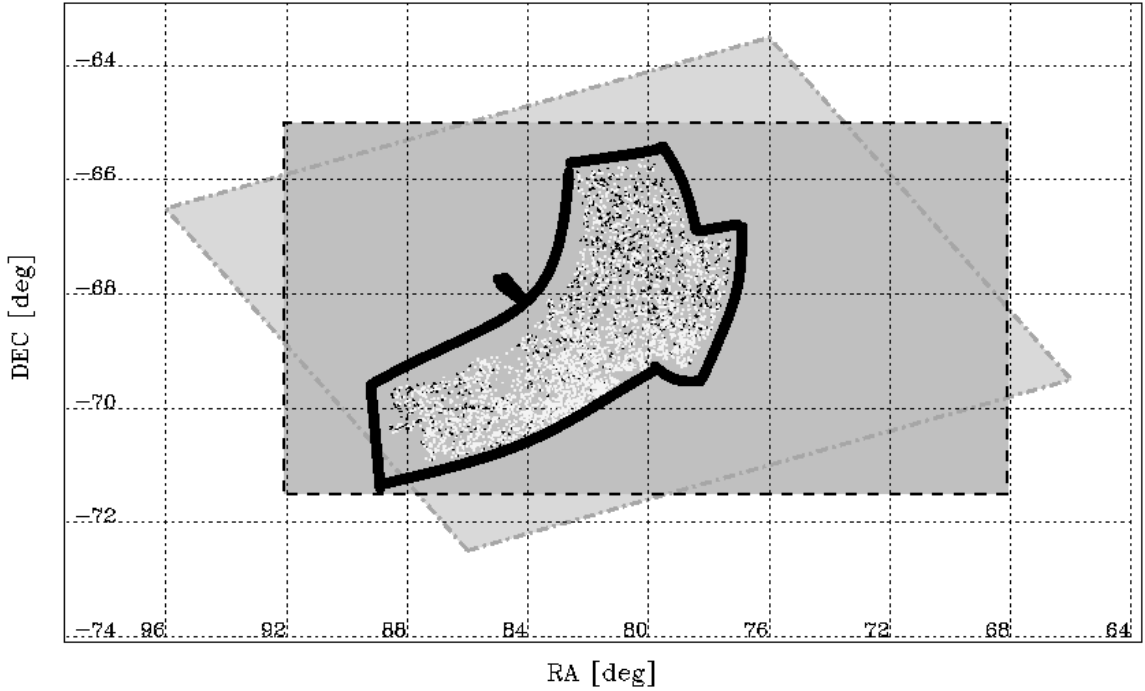}}
\caption{AKARI observed an area of about 10 $\rm deg^2$ of the Large Magellanic Cloud. The dash-dotted outline indicates the coverage 
of the Spitzer/SAGE survey [Meixner et al., 2006], and the dashed outline shows the coverage of the Large Magellanic Cloud optical photometric survey [Zaritsky et al., 2004]. 
AKARI sources not identified in any of selected databases
are shown with black triangles, and sources with identified counterparts by white squares.\label{fig:map}}
\end{figure*}

Observations in S11 and L15 data, which are unique to the AKARI survey, ensure new interesting features. AKARI data provided new information about mass-losing stars 
 [Kato et al., 2009; Onaka et al., 2009] and enabled studies of stars with circumstellar dust [Ita et al., 2009], the chemistry of the envelope (O- or C-rich), and the 
mass-loss rate accurately [Ita et al., 2008]. Ita et al. [2008, 2009] present new features on color-magnitude diagrams based on observations of an area of 10 $\rm deg^2$ 
of the Large Magellanic Cloud using the onboard AKARI Infrared Camera. Color-magnitude diagrams achieved from 11 and 15 $\mu$m data 
revealed new information about dust emission, enabled stars with circumstellar dust to be identified, and show that oxygen-rich, 
and carbon-rich, giants become blue, and turn red again during their evolution.

In this Work, we present the results from a multiwavelength catalog, based on the AKARI 3 $\mu$m survey. Our 
catalog contains data from a cross-correlation with a number of databases at different wavelengths. We show  
the separation of different classes of stars in the LMC 
in the color-magnitude, and color-color diagrams and analyze the contribution of different classes of sources at different infrared wavelengths to the total LMC flux, 
related to point sources.

This paper is organized as follows: in Sect.~\ref{r:data}, we show the basic information about the AKARI data: the sample with the complete 
color information. In Sections 3 and 4, 
we present and discuss how different types of objects separate in color-magnitude, and color-color, diagrams, respectively. In Sect.~\ref{r:flux}, we present the contribution 
from different objects to the total NIR and MIR flux in the LMC. We summarize in Sect.~\ref{r:summary}.

\section{The data}\label{r:data}

AKARI (ASTRO-F) is a Japanese infrared satellite launched on February 21, 2006 [Murakami et al., 2007]. 
AKARI has a lightweight telescope with a mirror of 68.5 cm effective diameter [Kawada et al., 2007], and two onboard 
instruments: an Infrared Camera (IRC) [Onaka et al., 2007] and a far-infrared surveyor (FIS) [Kawada et al., 2007]. 
Both instruments have a high-spatial-resolution/low- to-moderate-resolution spectroscopic capability. IRC has three
independent channels: NIR (1.8-5.5 $\mu$m), MIR-S (4.6-13.4 $\mu$m), and MIR-L (12.6-26.5 $\mu$m), nine imaging bands (three for
channel) and six dispersion elements (two for channel). 
NIR and MIR-S channels observed the same field of view due to the focal-plane layout, while MIR-L observed a field shifted by 25' with respect to the NIR/MIR-S center in a 
direction perpendicular to AKARI's orbit. 
Each channel has a wide field of view of about $10' \times 10'$, and adjacent fields overlap by $1.5'$. 
The photometry was independently performed for each of the pointed observations. Therefore, all sources that fall in the
overlapping regions between adjacent Fields of View (FOVs) have more than one photometric measurement. In those cases, only the source with the
best signal-to-noise (S/N) was listed in the catalog, and the other(s) were abandoned.

AKARI has
carried out two large-area legacy surveys: one of them was the LMC survey project (PI. T. Onaka) in the pointing mode. 
To map the LMC effectively, observations were performed during three separate seasons, from May 6, 2006, to June 8, 2006, from October 2, 2006, 
to December 31, 2006,  and from March 24, 2007, to July 2, 2007. 
As a positional reference, the Two Micron All-Sky Survey (2MASS) catalog [Skrutskie et al., 2006] was used.  At least five matched point sources were used to compute the coordinate
transform matrix that relates the image pixel coordinates to the sky coordinates. 
When sources were not found in 2MASS, transformation from the image pixel coordinates to the sky coordinates was based on matching detected sources with 
the SAGE point source catalog [Meixner et al., 2006]. The root-mean-squares of the residuals between the input 2MASS SAGE catalog coordinates, and the fitted coordinates, are 
smaller than 1: 2, 2: 6, and 2: 9, for NIR, MIR-S, and MIR-L images, respectively [Ita et al., 2008]. 

Bearing the above-mentioned observing strategy in mind, we should note that simultaneous observations of variable stars in different photometric bands are very important 
for constructing their correct infrared spectral energy distributions (SEDs) and for placing them correctly in color-magnitude, and color-color, diagrams. Due to intervals in observational time, 
the location of variable stars on color-color, and color-magnitude, diagrams will be biased by an additional scatter, because of changes in amplitudes and phases of variability. 
This effect has to be taken into account when interpreting color-color, and color-magnitude, diagrams of variable stars.


Observations of the LMC were carried out using IRC with wavelengths of N3 (3.2 $\mu$m), S7 (7.0 $\mu$m), S11 (11.0 $\mu$m), L15
(15.0 $\mu$m), L24 (24.0 $\mu$m), and a dispersion prism (2-5 $\mu$m, $\lambda/\delta\lambda \approx 20$). For the first time, observations were carried out with
imaging filters S11 and L15. These properties make the AKARI IRC unique, and complementary to the instruments onboard the 
Spitzer Space Telescope.
The NIR camera covers band N3, which can provide new information about carbon-rich objects. The AKARI IRC N3 filter is slightly bluer than the IRAC [3.6] filter, it covers 
the range from 
2.7 - 3.8 $\mu$m, while IRAC on board SST covers the range 3.15 - 3.90 $\mu$m. Ita et al. [2008] found new features of certain types of stars, comparing N3 and [3.6] photometries. 
Thanks to the higher sensitivity to the 3.1 $\mu$m HCN + $C_2H_2$ absorption of the N3 filter, than that of the [3.6] passband, observations in the N3 band are useful for identifying carbon-rich stars.
 
The entire LMC was also mapped by AKARI as part of the All-Sky survey at 6 bands in the mid- to far- infrared
wavelengths [Ishihara et al., 2006; Kawada et al., 2007] in 4 bands between 50 and 180 $\mu$m. Due to multiband
(11 bands) observations, AKARI provides a new window for studying star formation and galaxy evolution.
The LMC survey carried out over 600 pointing observations with IRC during three seasons from May 2006 to July 2007.


%


Observations over an area of about 10 $\rm deg^2$ of the LMC were performed in a pointing opportunity, in which the telescope was
pointed at a given position for about 10~min. There were detected over $5.9 \times 10^5$, $8.8 \times 10^4$, $6.4 \times 10^4$, $2.8 \times10^4$, and $1.5 \times10^4$
point sources at N3, S7, S11, L15, L24 bands, respectively [Ita et al., 2008, Kato et al., 2009]. In comparison, the Spitzer SAGE project detected only about 
$2 \times 10^5$ sources over an area of about $49~\rm deg^2$ of the LMC with all IRAC bands at two epochs separated by 3 months [Fazio et al., 2004]. 
The Release Candidate version 1 (hereafter RC1) of the AKARI LMC Large Area Survey point source catalog contains about a million sources [Ita, et al., 2008; Kato et al., 2012]. The fluxes for zero magnitude for 
the five IRC bands are 343.34, 74.956, 38.258, 16.034, and 8.0459 Jy [Tanab{\'e} et al., (2008)]. The 10-sigma limiting magnitudes are 16.8, 13.4, 11.5, 9.9 and 8.5 mag
at N3, S7, S11, L15, and L24, respectively, and the 90$\%$ completeness limits are 14.6, 13.4, 12.6, 10.7, and 9.3 mag at N3, S7, S11, L15, and L24, respectively 
(see Table~\ref{tab:summary_table}). 

%
%
%
%

\begin{table}[t]
 \renewcommand{\arraystretch}{0.70}
\vspace{-.3cm}
\caption{Survey properties.\label{tab:summary_table}}
\vspace{.1cm}
      \begin{threeparttable}

\begin{tabular}{llllll} \hline

Properties & N3 & S7 & S11 & L15 & L24\\ \hline \hline
Channel & NIR & MIR-S & MIR-S & MIR-L & MIR-L\\ \hline 
Reference wavelength [$\mu$m] & 3.2 & 7.0 & 11.0 & 15.0 & 24.0\\ \hline 
10 $\sigma$ detection limit [mag] & 16.8 & 13.4 & 11.5 & 9.9 & 8.5\\ \hline 
Saturation limit [mJy] & 250 & 1800 & 1800 & 2500 & 23000\\ \hline 
Saturation limit [mag]~\tnote{a} & 7.8 & 4.0 & 3.3 & 2.0 & -1.1\\ \hline 
Zero magnitude limits~\tnote{b} & 343.34 & 74.956 & 38.258 &  16.034 & 8.0459\\ \hline 
90 $\%$ completeness limits [mag] & 14.6 & 13.4 & 12.6 & 10.7 & 9.3\\ \hline 
Number of detected sources [Ita et al., 2008] & $> 5.9 \times 10^5$ & $> 8.8 \times 10^4$ & $> 6.4 \times 10^4$ &  $> 2.8 \times10^4$ & $> 1.5 \times10^4$\\ \hline 
Number of sources in RC1 -- Archive & $> 8.7 \times 10^5$ & $> 18.7 \times 10^4$ & $> 14.7 \times 10^4$ &  $> 11.1 \times10^4$ & $> 16.1 \times10^4$\\ \hline 
Number of sources in RC1 -- Catalog & $> 7.5 \times 10^5$ & $> 10.6 \times 10^4$ & $> 6.4 \times 10^4$ &  $> 2.8 \times10^4$ & $> 0.77 \times10^4$\\ \hline 
         \end{tabular}
         \begin{tablenotes}
           \item[a] Numbers are taken from ASTRO-F Observer’s Manual version 3.2.$^2$.
           \item[b] Tanab{\'e} et al., (2008).
         \end{tablenotes}
      \end{threeparttable}
\end{table}

RC1 includes two different sets of data: 
\begin{itemize}
 \item Catalogs -- including only reliable sources, i. e. sources matched with a Spitzer/SAGE source within a radius of 3 arcsec 
[Meixner et al., 2006].
 \item Archive -- including sources with a low reliability or large photometric uncertainties (all detected sources). 
\end{itemize}

As Archive should be used with caution, we based our first analysis on sources from Catalogs. This data set contains 754 288 sources, but only 7 722 were detected in 
L24 wavelength. 

For the first analysis, we selected from Catalogs all sources (3 852) detected at all NIR  and MIR wavelengths of: 3.2, 7.0, 11.0, 15.0, 24.0 $\mu$m, 
i. e. with a complete five-band color information. This sample of data was used to create color-magnitude, and color-color, diagrams presented in 
Sections 3 and 4, 
and an analysis of contributions from different objects to the total NIR and MIR flux in the LMC presented in Sect.~\ref{r:flux}.


In order to identify and classify sources, we searched for counterparts of selected sources based on the AKARI 3 $\mu$m survey in publicly-available databases at 
different wavelengths:
2MASS [Cutri et al., 2003; Skrutskie et al., 2003; Karachentseva et al., 2010], NED, SIMBAD, AKARI FIS All-Sky Survey [Yamamura et al., 2010], OGLE [Soszy{\'n}ski et al., 2008, b, 
2009a, b, c; Poleski et al., 2010a, b]. We cross-correlated our sources with these catalogs using
a positional tolerance of 3.0 arcsec, extended to 20 arcsec for the cross-correlation only with the AKARI-FIS data.  
If more than one object was found within the 
tolerance circle, the closest one was chosen. For almost 75$\%$ of the sample (2 836 from 3 852 sources), counterparts were found. 
In Fig.~\ref{fig:map}, a spatial distribution of identified and unidentified sources in the LMC is shown. 
Sources without a counterpart within a tolerance circle (1 016) should be further investigated. Such sources can be stars which are invisible due to a large amount of surrounding dust 
(e. g. OH/IR stars, dusty carbon stars, etc.).

\begin{figure}[t]
\centerline{\includegraphics[width=0.5\textwidth,clip]{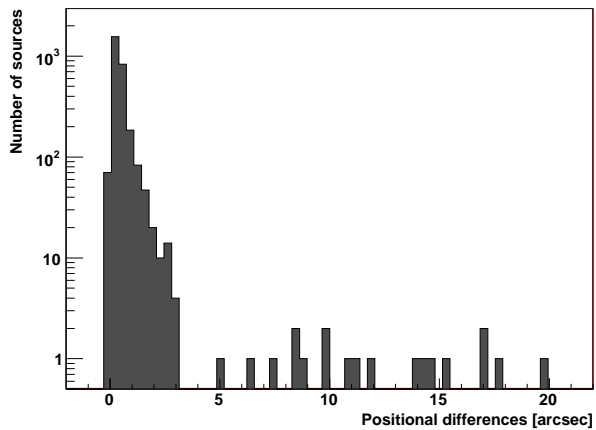}}
\caption{The distribution of positional differences between the sources and identified counterparts.The size of the bin is 0.3 arcsec.\label{fig:dist}}
\end{figure}

In Fig.~\ref{fig:dist}, a distribution of positional differences for matched sources (2 836) is shown. The mean positional difference is 0.57 arcsec, however a mean is not 
adequate for our sample due to an extension of positional tolerance to 20 arcsec for cross-correlation with the AKARI-FIS data. The median value of the positional difference 
is 0.41 arcsec. 
Figure A.1 shows a distribution of positional differences for matched sources only, with 2MASS, NED, SIMBAD, AKARI FIS All-Sky Survey, 
OGLE counterparts, respectively. 

We have found the largest amount of matches by cross-correlation of AKARI sources with 2MASS (2 661). Slightly fewer counterparts were identified in SIMBAD (2 233). With a 
tolerance radius of 3.0 arcsec, we have found 1 506 matches between AKARI sources and OGLE. By a cross-correlation with NED, we have identified only 244 objects. Although 
we have extended a positional difference to 20 arcsec for AKARI-FIS data only 77 sources were found. 
For about 35 $\%$ of sources (1 370), counterparts were found in all used databases excluding AKARI FIS All-Sky Survey and NED databases. 
Most of the counterparts were identified in 2MASS and SIMBAD, and for 1 389 sources the same object was identified in both databases. In some cases, we have identified sources only 
in one of the databases. 
We have found 525 matches in the tolerance radius only between AKARI sources and 2MASS. 117 AKARI sources were identified only in SIMBAD. For 9 sources, counterparts were 
found only in NED. With cross-correlation with AKARI-FIS data, we have found counterparts for 20 sources, and only for 2 with cross-matching solely with OGLE. 

For sources with identified counterparts (2 836), we compiled all the available information about types of objects from all the databases. 
Within our sample we have found 1 964 objects (51$\%$ of the sample) with a more detailed source type. If more than one object type was found for a single source 
we gathered all information, and if the types were different we have mostly followed the SIMBAD classification. We distinguish the following categories of objects:
\begin{itemize}
 \item AGB stars, including objects classified as AGBs, candidates for AGB, carbon stars, or Mira stars.
 \item other possible late-type pulsating giants (possible AGBs or similar), including objects classified as variable stars, stars suspected of variability, 
semiregular variables (SRVs), or OGLE small amplitude red giants (OSARGs). 
 \item post-AGB stars, including objects classified as Planetary Nebula or possible Planetary Nebula.
 \item Young stellar objects, including objects classified as YSOs or candidates for YSO.
 \item multiple systems, including objects classified as stars in a cluster, stars in association, multiple objects, clusters of stars or double or multiple stars.
 \item sources of unknown origin, including objects classified as Radio-/ IR-/ UV-/ X-/Red- sources.
 \item background objects, classified as galaxies, quasi-stellar objects (QSOs) or active galactic nucleus (AGNs).
 \item foreground objects, classified as objects known to be placed within the Milky Way Galaxy.
 \item others, including one of the remaining objects with an identified type but not included in the above categories. 
\end{itemize}

\begin{figure*}[t]
\centerline{\includegraphics[angle=90, width=0.95\textwidth,clip]{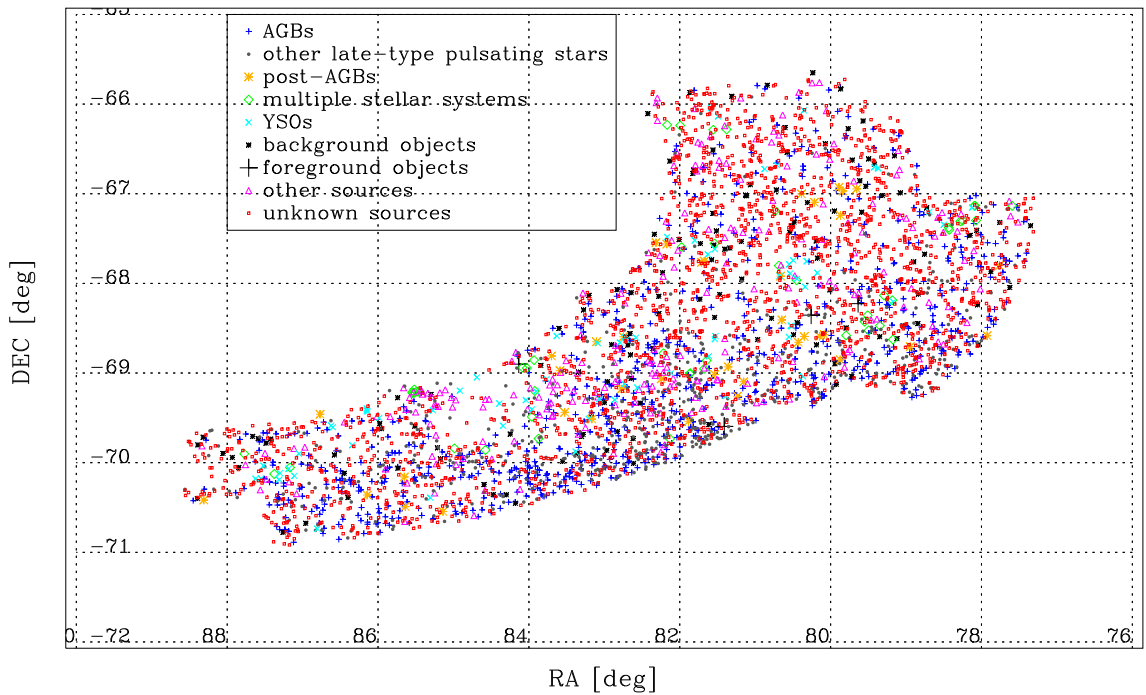}}
\caption{Spatial distribution of different sources. AGB stars are shown as small plus signs, other late-type pulsating giants as gray filled circles, post-AGB stars as asterisks, 
multiple stellar systems as gray diamonds, YSOs as gray X signs, 
background objects as black asterisks, foreground objects as gray plus signs, other sources as triangles, and unknown sources small, open squares.\label{fig:maps_types}}
\end{figure*}

In Fig.~\ref{fig:maps_types}, a spatial distribution of different categories of objects in the Large Magellanic Cloud is shown.

Among the identified sources, candidates for AGB stars dominate and amount to 42$\%$ of the objects of identified types. If classification is tentative, they are classified 
as other late-type pulsating giants (27$\%$ of the sample), which means that if these objects are not exactly AGBs, they are pulsating red giants not so far from AGBs in their 
evolutionary and physical properties. More than 6.7$\%$ of the selected group are background 
objects, mainly galaxies (about 6.5$\%$). We also found a group of young stellar objects (3.6$\%$), multiple systems (2.5$\%$), post-AGBs (2.0$\%$) and others of which 
mostly are related to stars. Within the sample, only 4 objects identified as Galactic sources were found, however in our sample there are probably more Milky Way objects.
  
The effect of unequal Galactic foreground extinction, as well as that of the local interstellar extinction, due to the LMC disk material, 
should also be analyzed carefully, particularly for N3 band data. Based on Sakon et al. [2006], the Galactic foreground emission at COBE DIRBE bands range from 3.6 to 7.5 MJy/sr at 100 $\mu$m and 
from 6.4 to 14.7 MJy/sr at 140 $\mu$m, assuming a correlation between the HI and infrared emissions within the AKARI LMC Large Area survey area, where the Galactic HI 
intensity ranges from 300 - 600 K km/s. The optical depth at 100 $\mu$m can be estimated as $\tau$(100 $\mu$m)=2.4$\cdot10^{-4}$ -7.2$\cdot10^{-4}$. In this case, the effect of Galactic 
foreground extinction variations on the N3 band strength of sources within the observed area must be smaller than a few percent even if $A_{N3}/\tau(100\mu m)\approx30$ is assumed 
[Sakon et al. 2006]. 

A discussion of foreground and background objects is presented in Sect.~\ref{r:cmdforeback}. The fractions of different types of objects among IR-bright LMC sources are 
shown in Table B.1.
The properties of AGB, other late-type pulsating giants and post-AGB stars are summarized in Table C.1.

There remains a possibility of missidentification, since the gathered data are compilations of 
data from several databases. However, a few possible 
missidentifications should not affect the final result of this Work.

\section{Color-magnitude diagrams}\label{r:cmd}

Color-magnitude diagrams are a tool often used in astronomy and often show a clear separation of different objects. 
Infrared color-magnitude diagrams have been often 
used for the analysis of Spitzer/SAGE data [Meixner et al., 2006; Blum et al., 2006; Bonanos et al., 2009].
Adding S11 and L15 data (unique to the AKARI survey) ensures additional new features
and can be essential for the estimation of the silicate band strength in mass-losing stars [Kato et al., 2009; Onaka et al., 2009]. 
Color-magnitude diagrams have been used to identify stars with circumstellar dust and to show that 
dust emission is related, not only to sources brighter than the tip of the first red giant branch, but also to fainter 
red giants [Ita et al., 2009]. 
Ita et al. [2009] correlates this new sequence with red giants with aluminum oxide dust without the silicate feature. 
IRC limits enable detection only to the brightest end of the YSOs- stars such as T Tauri are too deep for AKARI IRC 
survey - although young stars of intermediate mass 
(Herbig Ae/Be), or classical Be stars, can be almost completely detected. 
IRC is capable of detecting all mass-losing AGB stars [Ita et al., 2008]. 
The N3 band is sensitive to the 3.1 $\mu$m $\rm HCN+C_2H_2$ absorption feature. 
This absorption line is a characteristic of carbon-rich AGB stars [e.g, van Loon et al., 1999] but  
can also be atributed to water ice [Manteiga et al., 2011]. 
On the basis of the N3 color, it is possible to determine which sources are carbon-rich. 
Observations in the S11, L15, L24 bands can ensure information about the chemistry of the envelope 
(O- or C-rich), and the mass-loss rate accurately, thanks to the presence of silicate bands at 
10 and 18 $\mu$m, and the silicon carbide band at 11.3 $\mu$m [Ita et al., 2008].
  
\begin{figure}[t]
\centerline{\includegraphics[width=0.25\textwidth,clip]{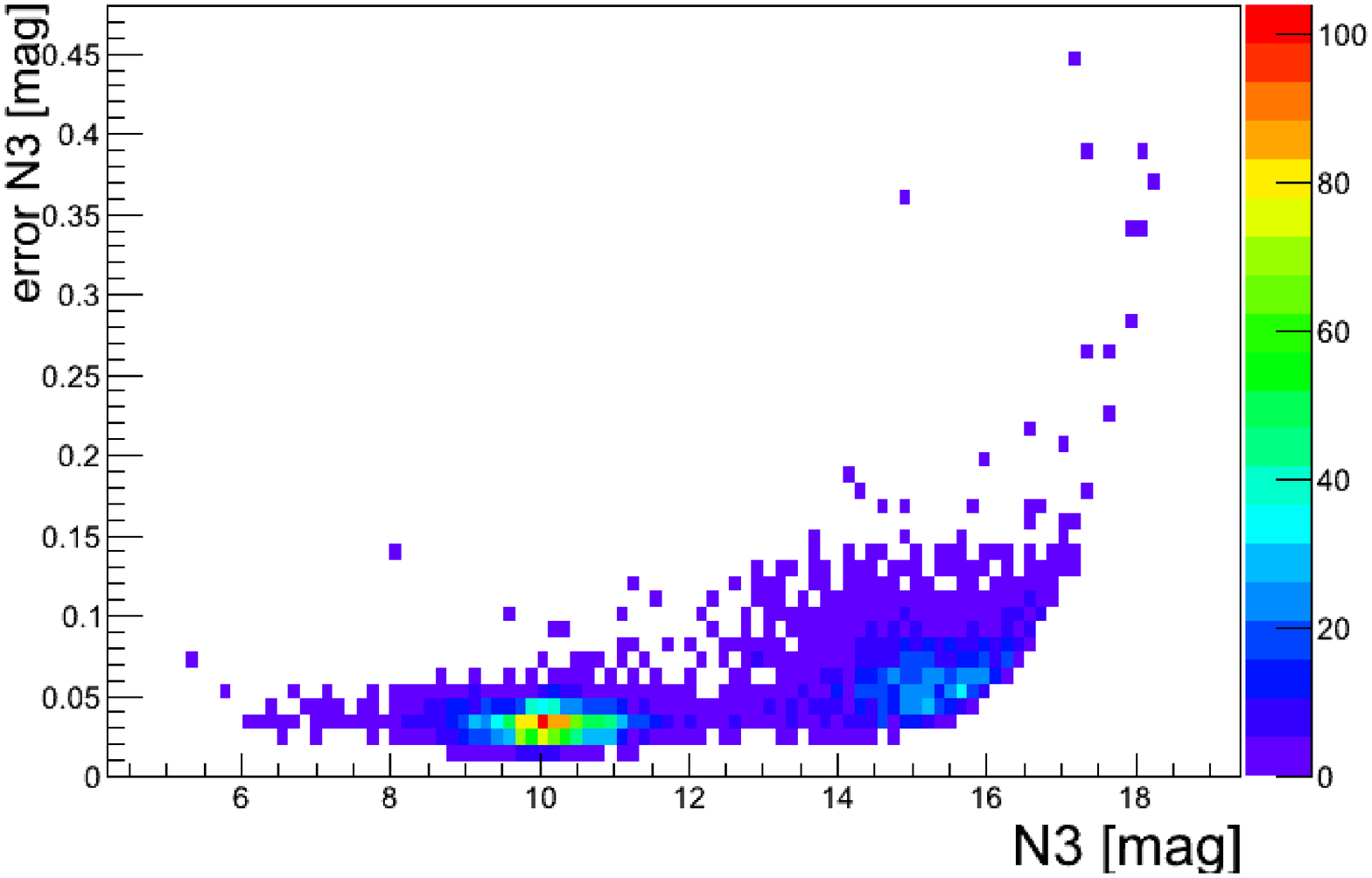}
\includegraphics[width=0.25\textwidth,clip]{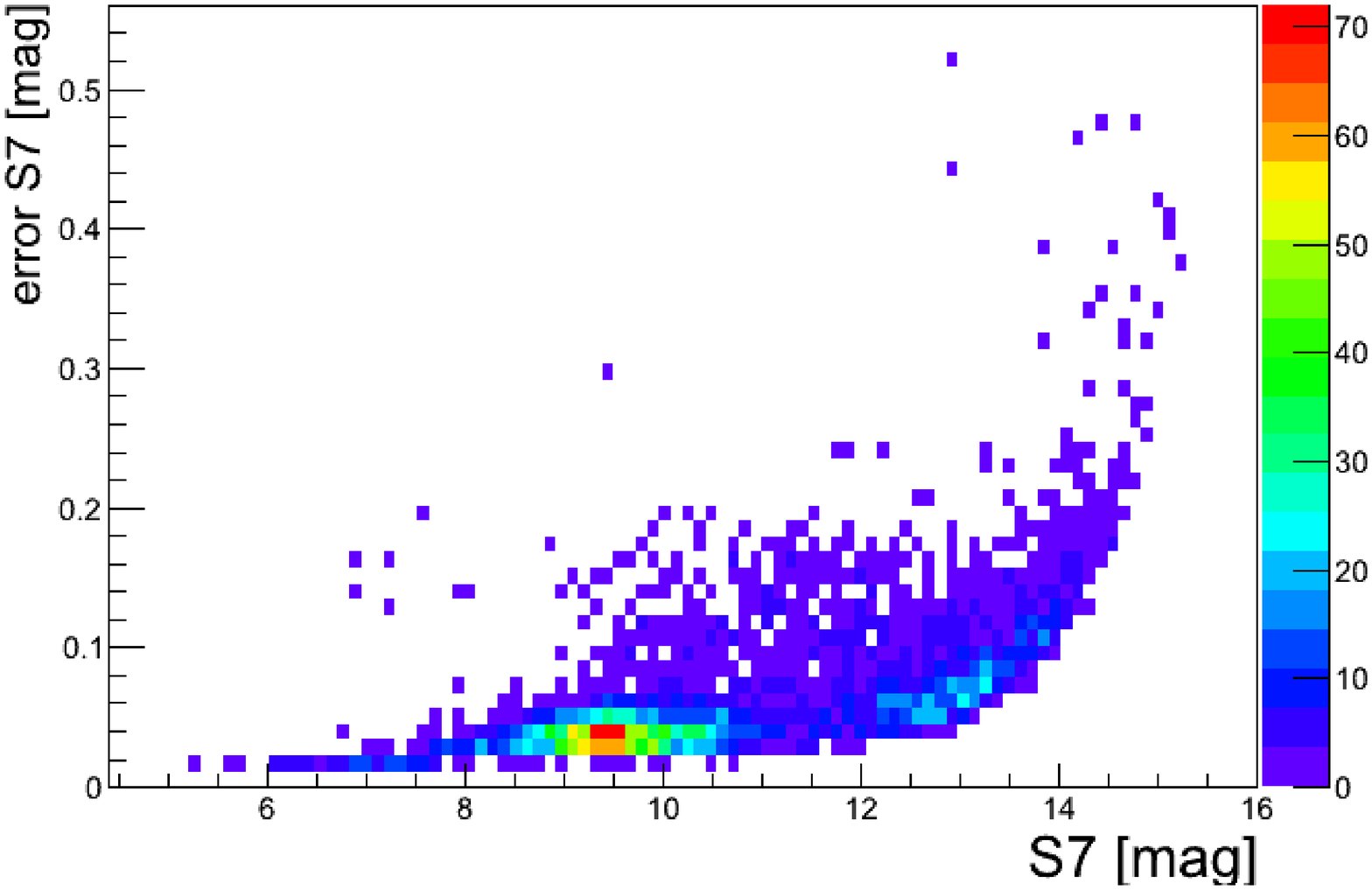}}
\centerline{\includegraphics[width=0.25\textwidth,clip]{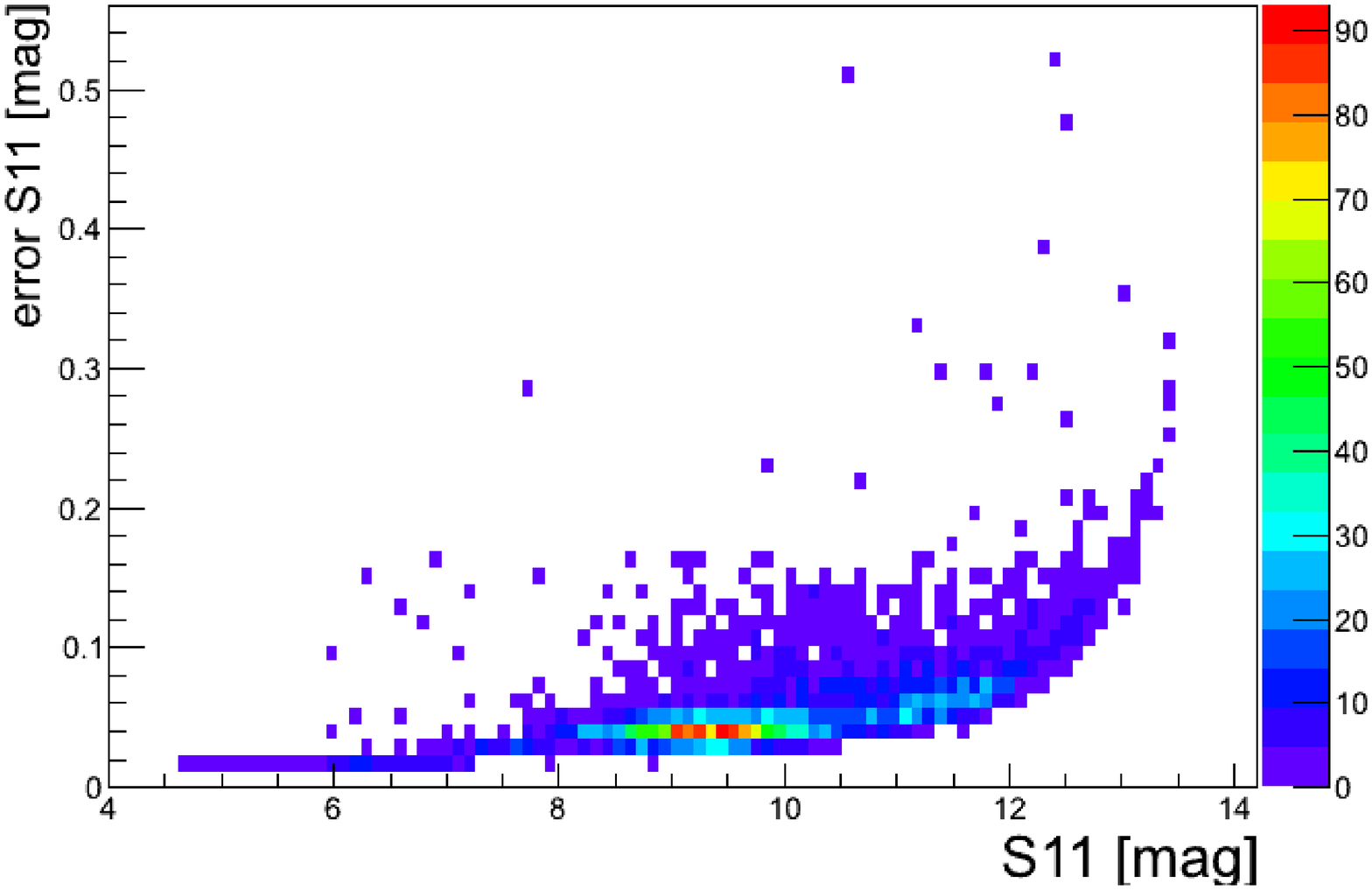}
\includegraphics[width=0.25\textwidth,clip]{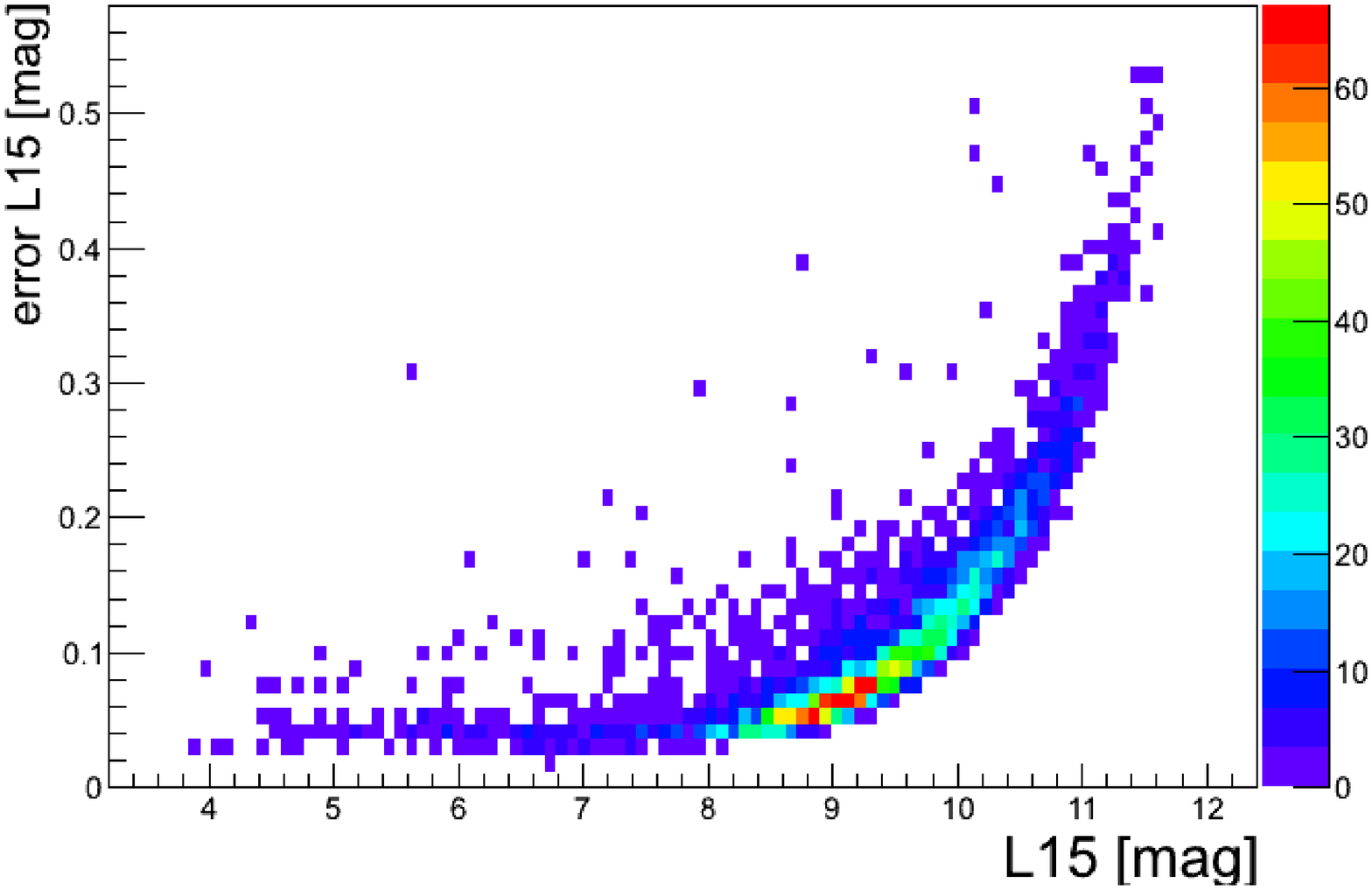}}
\centerline{\includegraphics[width=0.25\textwidth,clip]{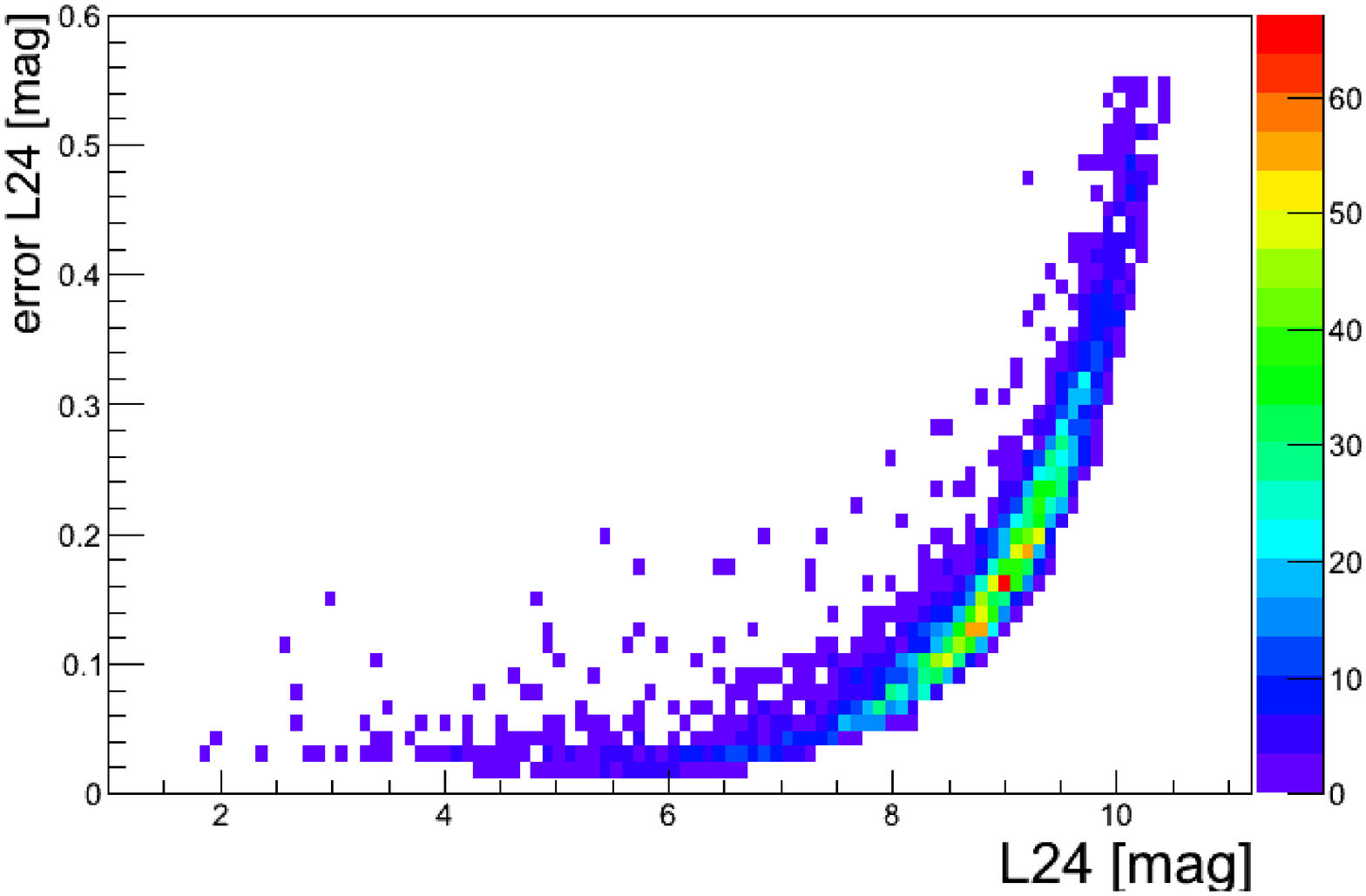}}
\caption{Photometric uncertainties as a function of the magnitude at
each IRC band for sources with full color information.\label{fig:error}}
\end{figure}

In this section, 
we analyze color-magnitude diagrams with all different types of sources (Sect.~\ref{r:cmdall}), 
and we also make further comparison of only AGB and post-AGB stars 
(Sect.~\ref{r:cmdagb}). 
The distributions of the photometric uncertainty versus the magnitude in the IRC bands for selected sources are shown in Fig.~\ref{fig:error}. More about the accuracy of the photometry for all the samples 
can be found in Ita et al., [2008]. More information about the IRC imaging data pipeline can be found in the IRC Data User's
Manual~\footnote{$^{2}$ http://www.ir.isas.jaxa.jp/ASTRO-F/Observation/\#IDUM}. 
Photometric uncertainty for our sample remains negligible when we analyze color-color, and color-magnitude, diagrams. 
However, differences in phase may play a role in creating a dispersion of the properties 
of variable stars in these plots. 

\subsection{Separation of different types of sources in color-magnitude diagrams}\label{r:cmdall}
\begin{figure}[t]
\centerline{\includegraphics[width=0.411\textwidth,clip]{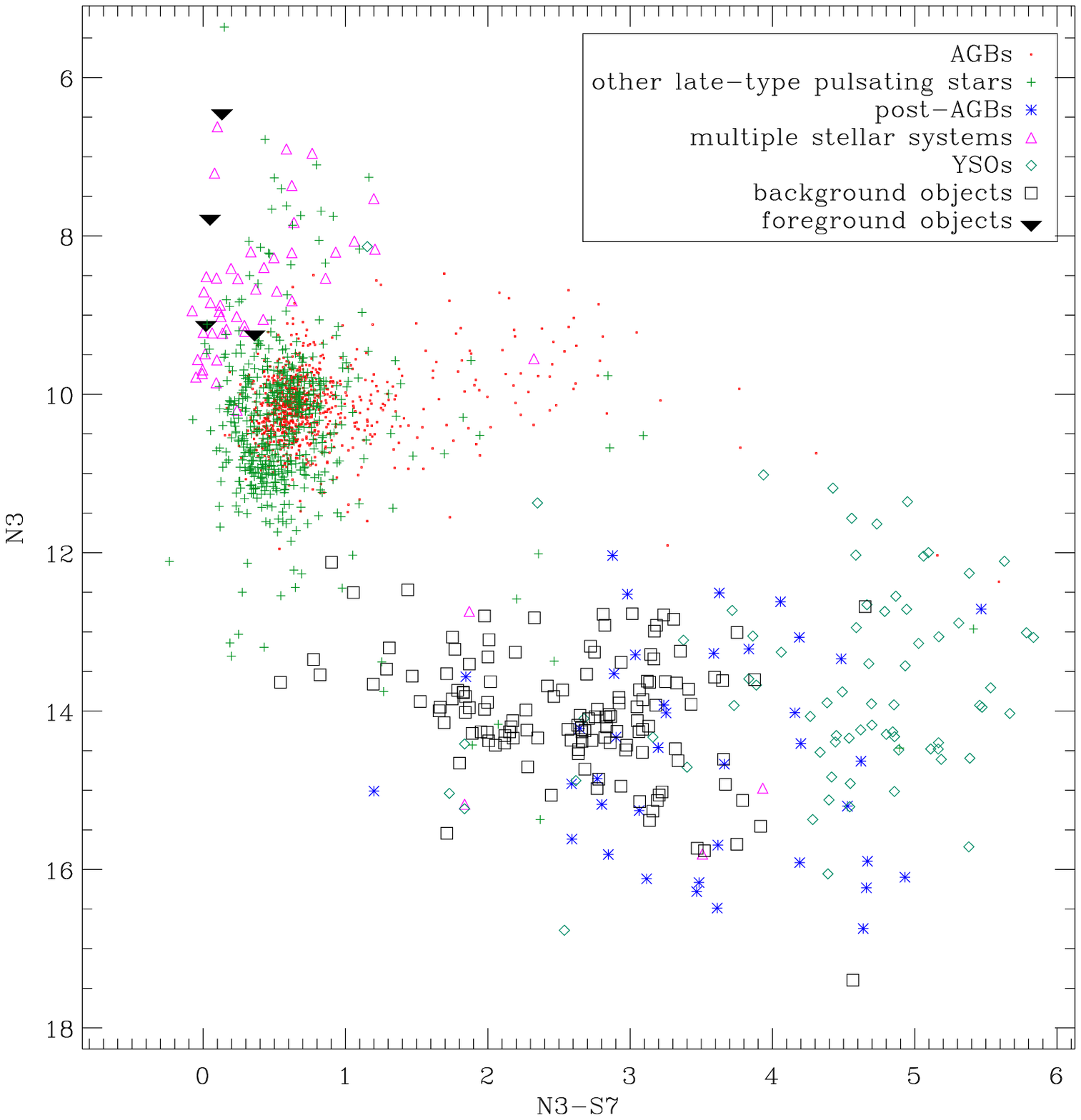}}
\centerline{\includegraphics[width=0.411\textwidth,clip]{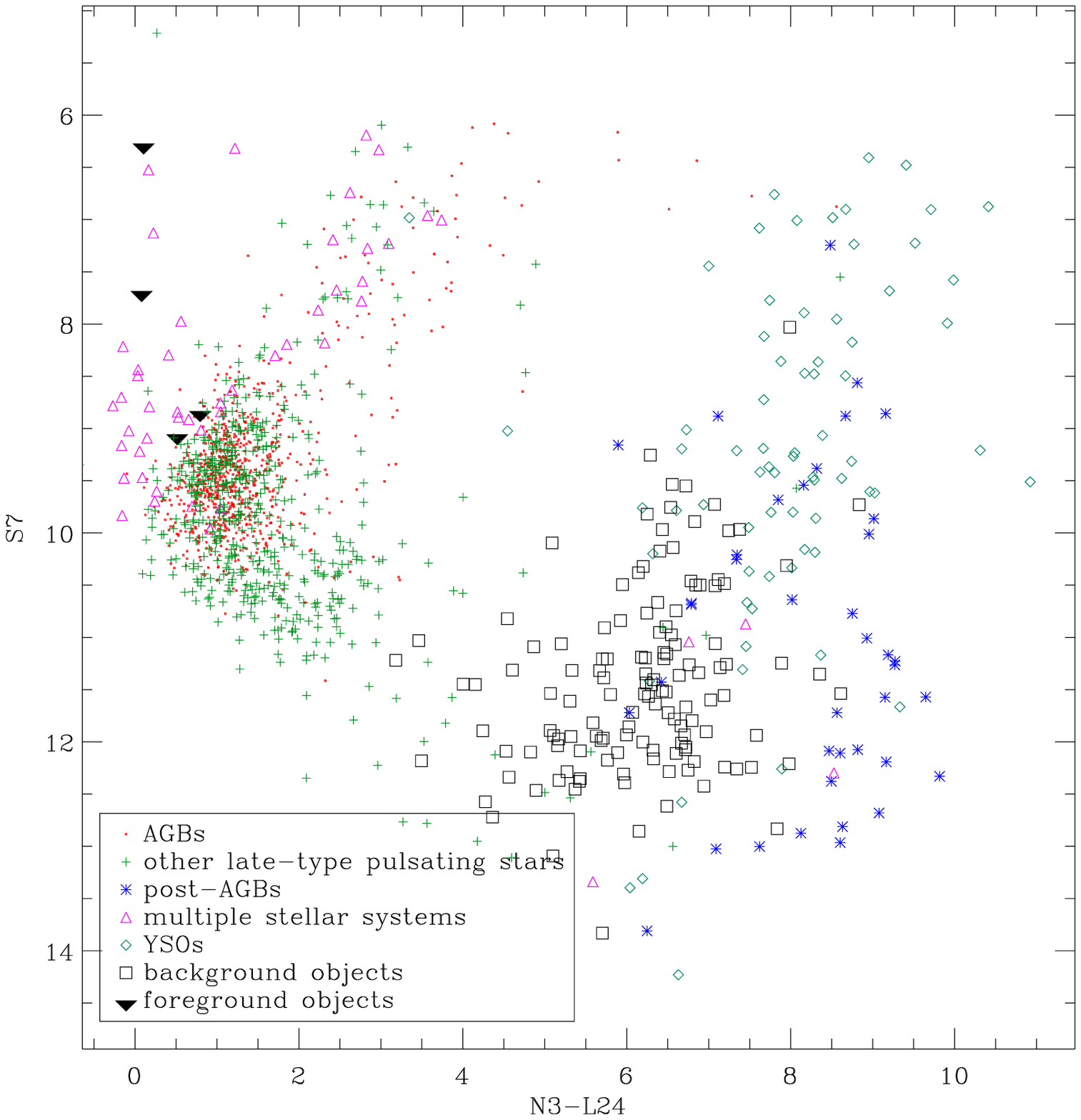}}
\centerline{\includegraphics[width=0.411\textwidth,clip]{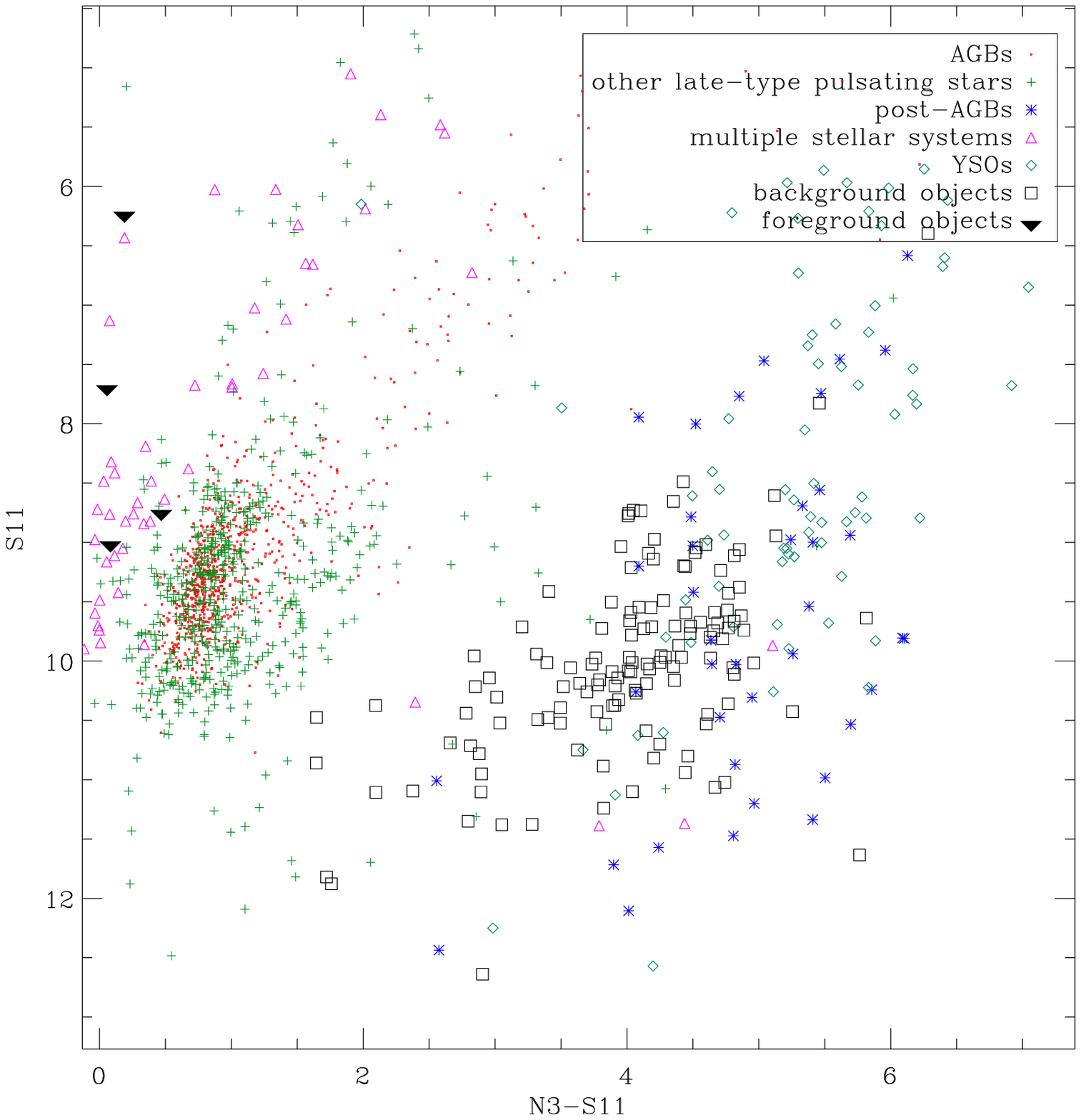}}
\caption{Position of different types of objects in the color-magnitude diagrams $N3-S7$ vs. 
N3 (upper), $N3-L24$ vs. S7 (middle), $N3-S11$ vs. S11 (lower). 
AGB stars are shown as small gray filled circles, other late-type pulsating giants as gray plus signs, post-AGB stars as asterisks, 
multiple stellar systems as gray open triangles, YSOs as gray diamonds, 
background objects as black open squares, and foreground objects as black filled triangles.\label{fig:cmd}}
\end{figure}

In order to analyze color-magnitude diagrams, we discriminated among our sample following groups of objects: \\
(1) AGB stars (including sources classified as : AGBs, candidates for AGBs, carbon stars, or Mira stars); \\
(2) other possible late-type pulsating giants; \\
(3) post-AGB stars, mainly Planetary Nebulae and protoplanetary nebulae;\\
(4) multiple stellar systems;\\
(5) YSOs; \\
(6) background objects; and \\
(7) foreground objects. \\
Color-magnitude diagrams of different types of sources from our sample are shown 
in Fig.~\ref{fig:cmd}. 
In all the color-magnitude diagrams different types of objects are well separated, there is an especially prominent 
separation between a group consisting of multiple stellar systems, 
possible late-type pulsating giants, AGB stars, and foreground objects, and a group including YSOs, 
post-AGB stars and background objects, as is clearly shown in 
Fig.~\ref{fig:cmd}. 
As shown in  diagram $N3-S7$ vs. N3 (upper panel of 
Fig.~\ref{fig:cmd}) post-AGB stars, YSOs, and background objects, are among the least 
luminous stars at 3.2 $\mu$m. 
The most luminous groups are multiple stellar systems. 
Also a group of other late-type giants tend to be among the most luminous stars at 3.2 
$\mu$m, but this group is nonuniform and objects are spread over the whole diagram. 
AGB stars are less luminous than multiple stellar systems, but they are significantly 
more luminous than YSOs and post-AGB stars. 
AGB stars, other late-type pulsating giants, multiple stellar systems, and foreground 
objects, occupy the ``bluer'' part of the color-magnitude diagrams and YSOs, 
post-AGB stars, and background objects, tend to be found in the ``redder'' part. 
YSOs tend to occupy the ``reddest'' part of the diagrams, while post-AGB stars 
and background objects are scattered. 
AGB stars form a large cloud in the ``bluer'' part of the diagram, with a long tail going 
into the ``redder'' part. 
Due to a small amount of foreground objects, they are not well separated but they are 
placed in the ``bluer'' regions and are similar to, or more luminous than, AGB stars. 

Post-AGB stars and background objects also tend to be the 
least luminous at 7.0 $\mu$m, which is visible in the diagram $N3-L24$ vs. S7 
( middle panel of Fig.~\ref{fig:cmd}), but the luminosity of YSOs is comparable to 
the luminosity of AGB-stars. 
Also, multiple objects have a similar tendency as in 3.2 $\mu$m and they are among 
the most luminous stars, but, at 7.0 $\mu$m some AGB stars also tend 
to be among the most luminous objects. 
YSOs and post-AGBs are ``redder'' than AGB stars and possible AGBs or similar giants. 
At 11 $\mu$m it is not clear what kind of objects are the most luminous because each 
group spans quite a wide range of luminosity. In the $N3-S11$ vs. S11 
diagram (lower panel of Fig.~\ref{fig:cmd}) a tendency to form clouds by all groups 
stretching from a lower to a higher luminosity is visible. Most of the multiple objects and AGBs 
represent luminosities similar to the background objects, which are less luminous than YSOs. 
Yet, AGB stars, pulsating giants and multiple systems are well separated from post-AGB stars, 
YSOs and background objects. 
Multiple stellar systems, foreground objects, and AGB stars, occupy the ``bluer'' part of 
the diagram, while YSOs, post-AGB stars, and background 
objects are found in the ``redder'' part. 
At 15.0 and 24.0 $\mu$m, all different types of objects are widely spread over the luminosity 
range and it is hard to distinguish the most luminous group.   

\subsection{Separation of AGB and post-AGB stars in color-magnitude diagrams}\label{r:cmdagb}

\begin{figure}[t]
\centerline{\includegraphics[width=0.411\textwidth,clip]{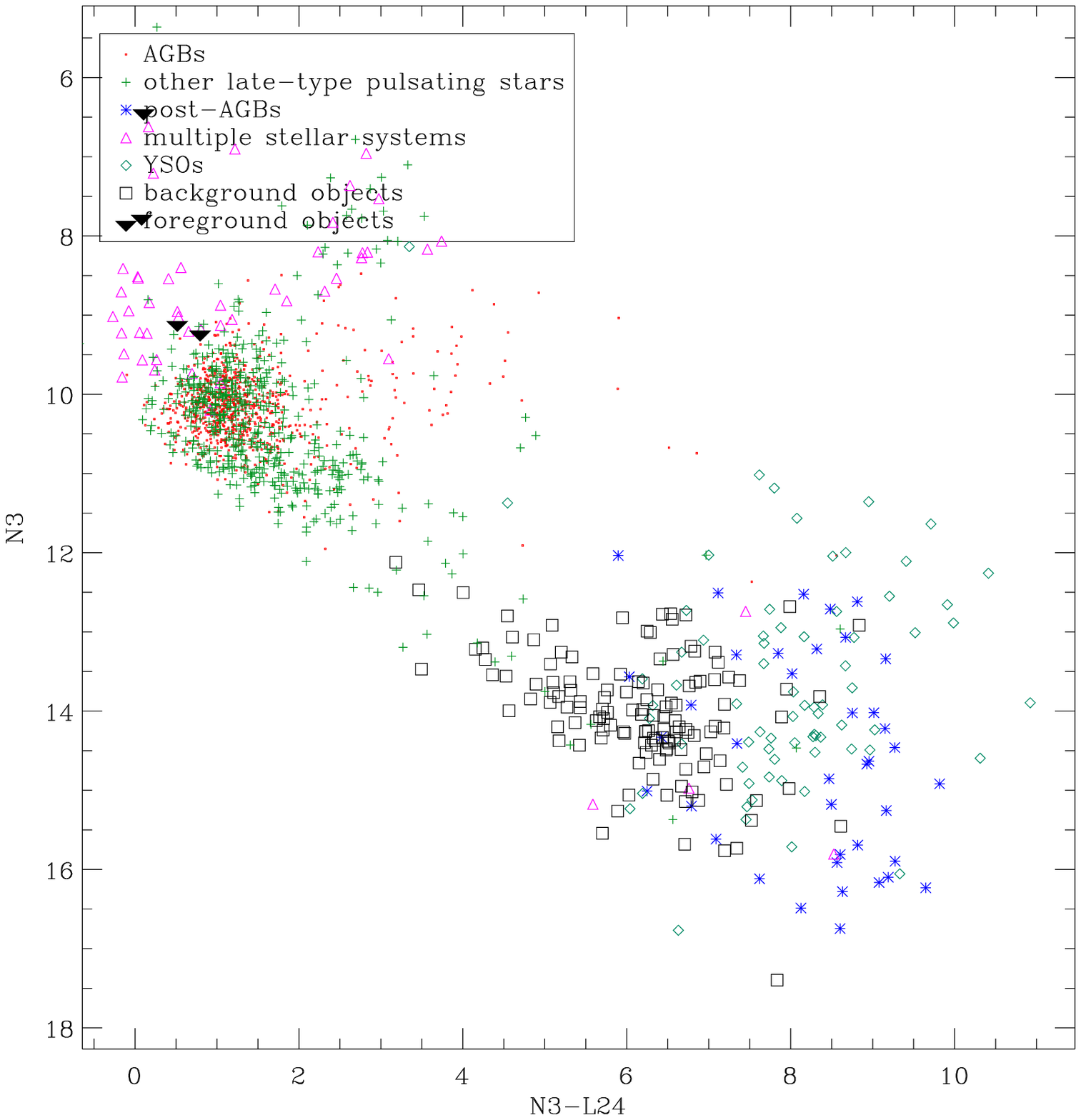}}
\centerline{\includegraphics[width=0.411\textwidth,clip]{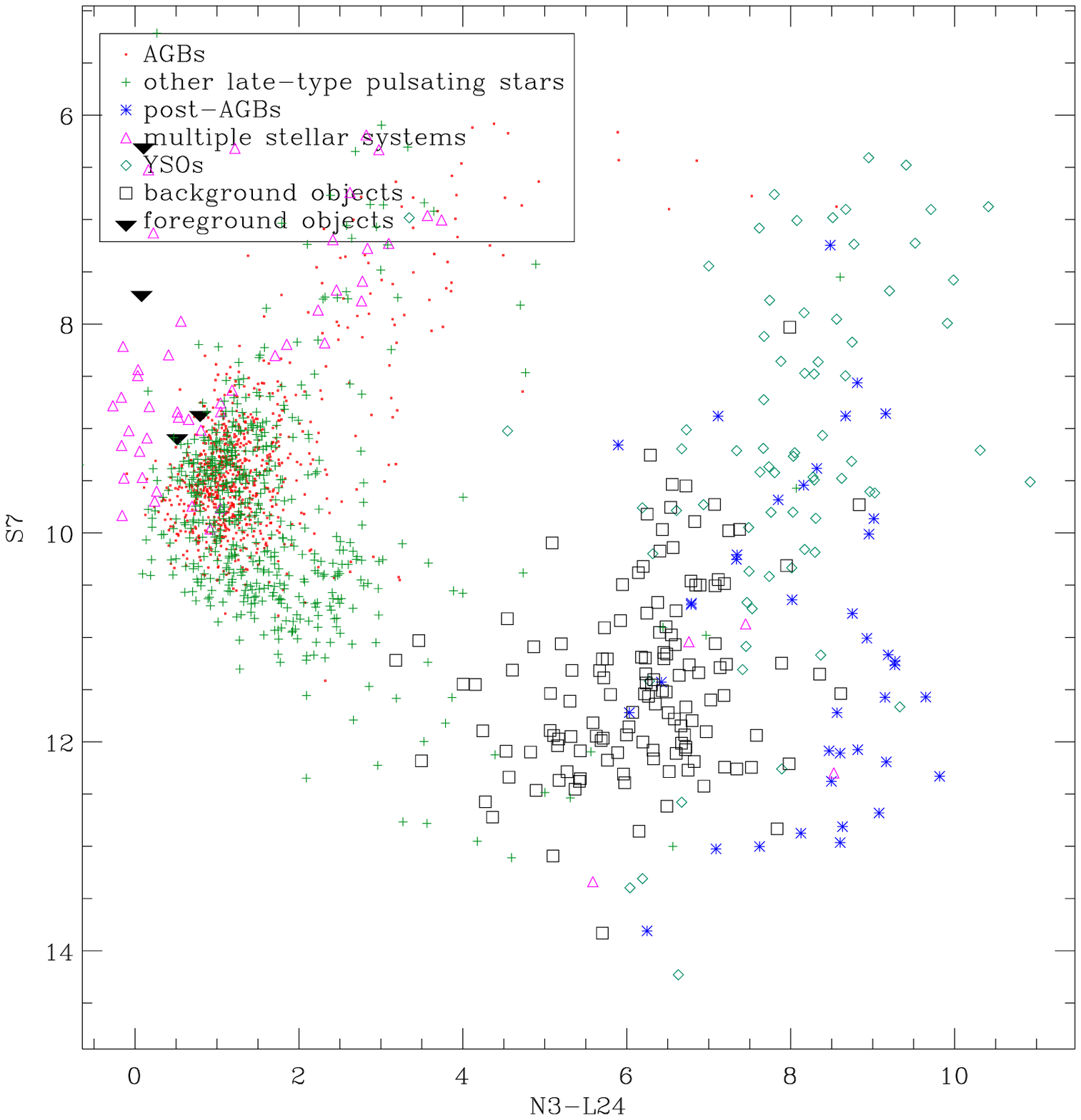}}
\centerline{\includegraphics[width=0.411\textwidth,clip]{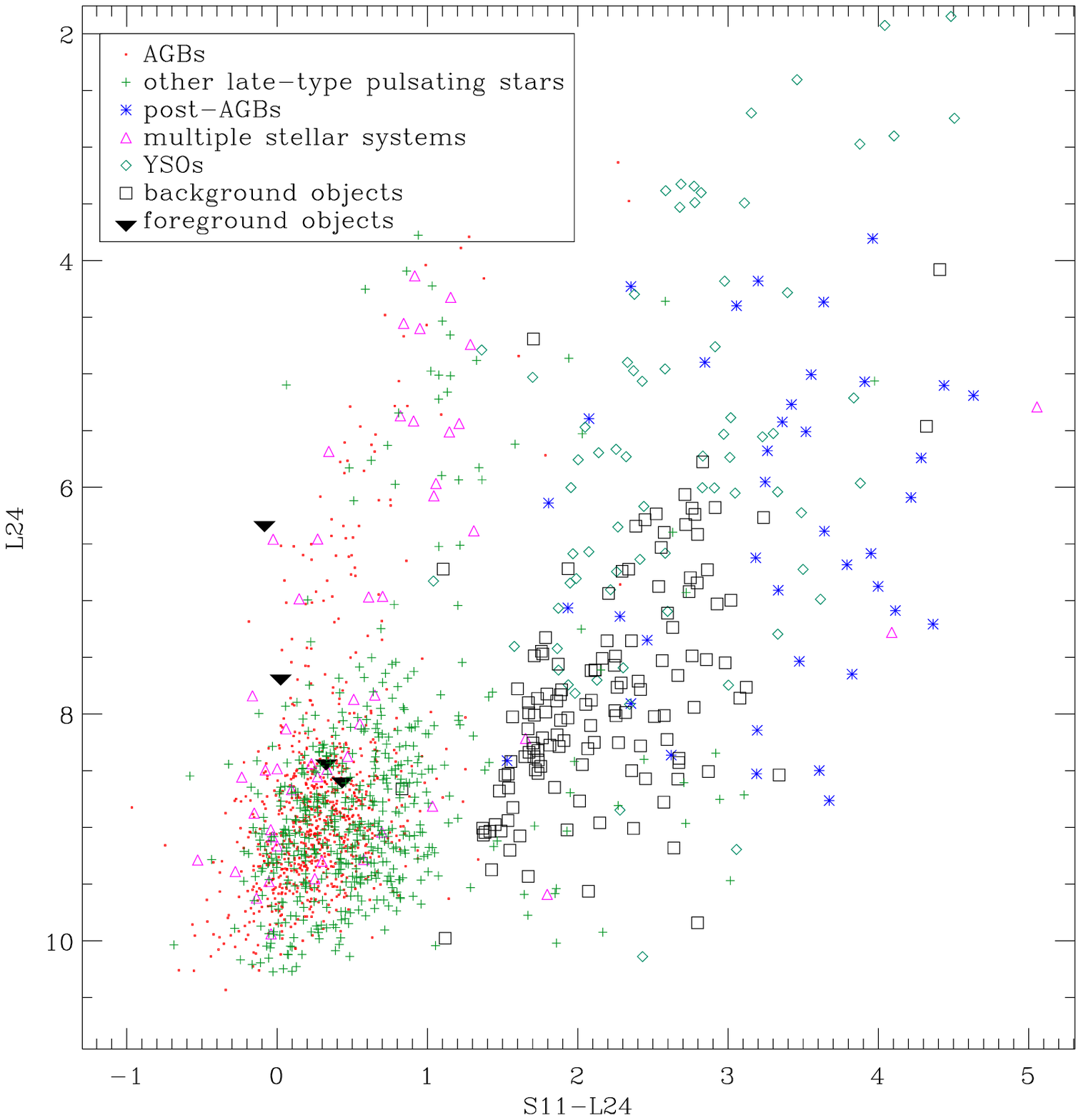}}
\caption{Position of different types of objects in the color-magnitude diagrams $N3-L24$ vs. N3 (upper), 
$N3-L24$ vs. S7 (middle), $S11-L24$ vs. L24 (lower). 
(d) AGB candidates  are shown as plus signs, (a) post-AGB stars as gray asterisks, 
(b) Mira stars as gray open triangles, (c) carbon stars as black filled triangles, (e) variable stars as X signs, 
(f) SRVs as gray open squares, and (g) OSARGs as diamonds.\label{fig:cmdagb}}
\end{figure}

In this section, 
we discuss the color-magnitude diagrams of AGBs, other late-type pulsating stars possibly being AGBs, or close to the AGB phase, and post-AGBs (groups (1), (2), and (3) 
from~\ref{r:cmdall}). 
We divide the sources into the following types: \\
(a) post-AGB stars (including Planetary Nebulae); \\
(b) Mira stars; \\
(c) carbon stars; \\
(d) candidates for AGB stars (including sources which probably are AGB stars but their type was not confirmed); \\
(e) variable stars; \\
(f) semiregular variables (SRVs); and \\
(g) OGLE small amplitude red giants (OSARGs). \\
Color-magnitude diagrams for the AGB group are presented in Fig.~\ref{fig:cmdagb}.  
In all color-magnitude diagrams, different types are well separated. Post-AGB star, especially, form 
a separate group in the ``redder'' regions of the diagrams. 
Also, Mira stars tend to form a compact, clear separated cloud in the ``bluer'' part of the diagrams, 
especially in diagram $N3-L24$ vs. N3 (upper panel of Fig.~\ref{fig:cmdagb}), and $S11-L24$ vs. L24 
(lower panel of Fig.~\ref{fig:cmdagb}). 
As shown in  diagram  $N3-L24$ vs. N3 (upper panel of Fig.~\ref{fig:cmdagb}), a group of (e) variable stars, (f) SRVs, 
 and (g) OSARGs, are among the most luminous stars at 3.2 $\mu$m, and post-AGB stars 
are among the least luminous objects. 
Variable stars do not seem to form any separate cloud, they are spread over the whole diagram. 
As shown in  diagram  $N3-L24$ vs. S7 (middle panel of Fig.~\ref{fig:cmdagb}), SRVs, a small group 
of variable stars, are among the most luminous stars, not only at 3.2$\mu$m, but also at 7 $\mu$m, 
but also in this band Mira stars tend to have similar luminosities. 
Post-AGB stars, as at 3.2$\mu$m, are also among the least luminous objects at 7 $\mu$m, and 
they occupy the ``redder'' part of the diagram.
However, post-AGB stars tend to be more luminous than carbon stars and AGB stars at 24 $\mu$m, 
although a group of post-AGBs are quite widely spread in luminosity.

Color-magnitude diagrams presented by Ita et al. [2008] show a clear separation in luminosity 
and color between carbon stars, dusty red giants and YSO candidates. 
Since Ita et al. [2008] estimate the absolute magnitude by adding the distance modulus to the observed magnitude, and we do not, 
a straightforward comparison of our diagrams with those of Ita et al [2008] is not possible.  
 
Moreover, our sample is different. Ita et al. [2008] based their analysis on all detected sources, i.e. over $5.9 \times 10^5$, $8.8 \times 10^4$, $6.4 \times 10^4$, $2.8 
\times10^4$, and $1.5 \times10^4$, point sources at the N3, S7, S11, L15, L24 bands, respectively. Our work is based only on reliable data provided by RC1 with full color information. 
A comparison of the number of sources used by Ita et al. [2008], and in this paper, is presented in Table~\ref{tab:summary_table}.

YSO candidates have redder colors than other groups, both in our, and the Ita et al. [2008], 
color-magnitude diagrams. Also, color-magnitude diagrams with Spitzer data show that YSOs are among the reddest objects [Meixner et al., 2006]. 
In the mid-infrared diagrams presented by Ita et al. [2008], dusty red giants are brighter than optical 
carbon stars. As we have divided our sample into different categories of types it is not easy to verify this 
information.  
However, in our infrared color-magnitude diagrams, Mira stars are more luminous than carbon stars. 
Observations in the N3 band, thanks to the sensitivity to 3.1 $\mu$m HCN + C$_2$H$_2$, could be useful for separating 
carbon-rich AGBs on the basis of their colors. 

\section{Color-color diagrams}\label{r:color}
As it was discussed in Sect.~\ref{r:cmd}, color-magnitude diagrams can be used to distinguish different types 
of objects. 
Also color-color diagrams give us useful information about the properties of different objects. 
As in the previous Section, we analyze color-color diagrams with all different types of sources 
(Sect.~\ref{r:colorall}). 
In Sect.~\ref{r:coloragb}, we present a further comparison of the group of AGB and post-AGB stars.

\subsection{Separation of different types of sources in color-color diagrams}\label{r:colorall}

\begin{figure}[t]
\centerline{\includegraphics[width=0.411\textwidth,clip]{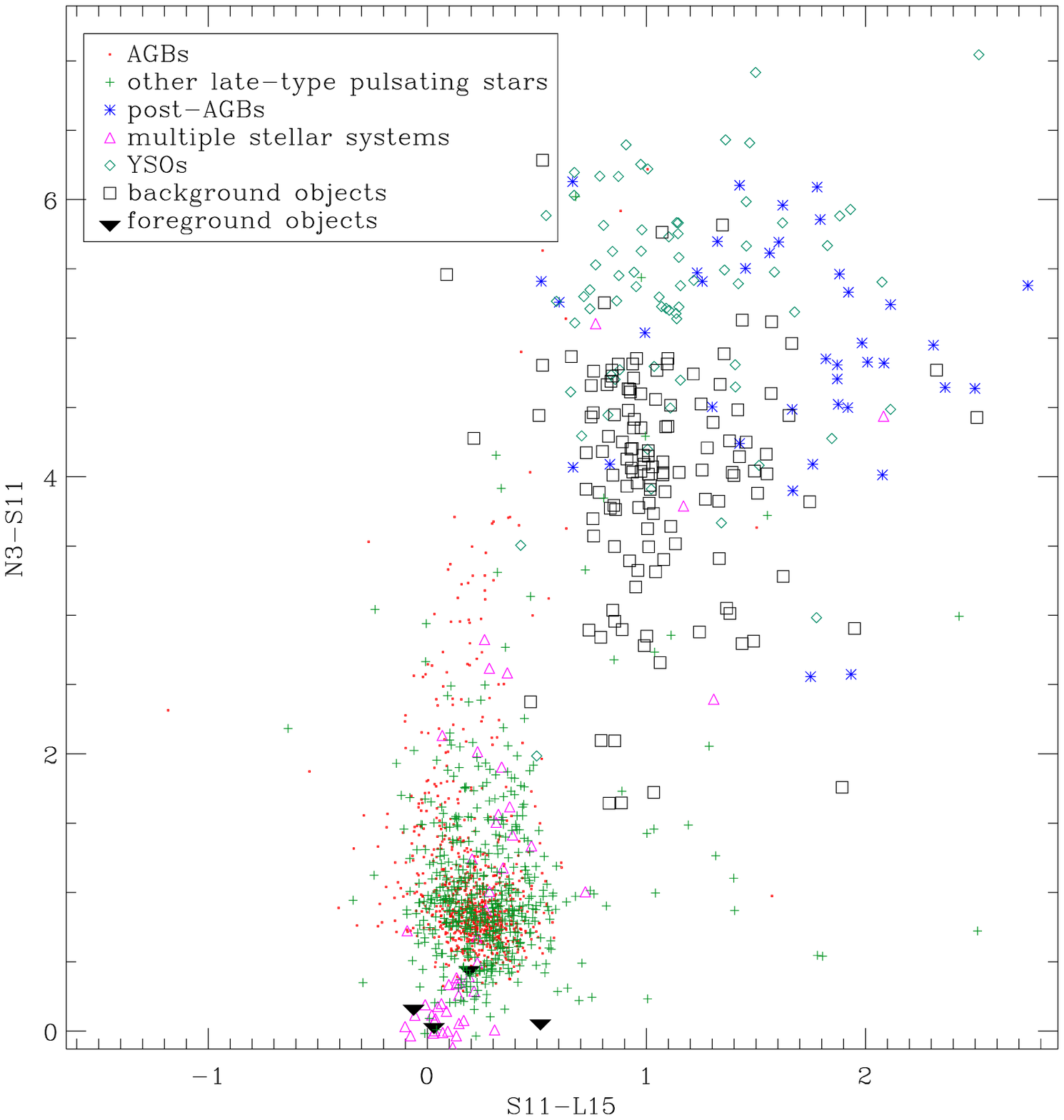}}
\centerline{\includegraphics[width=0.411\textwidth,clip]{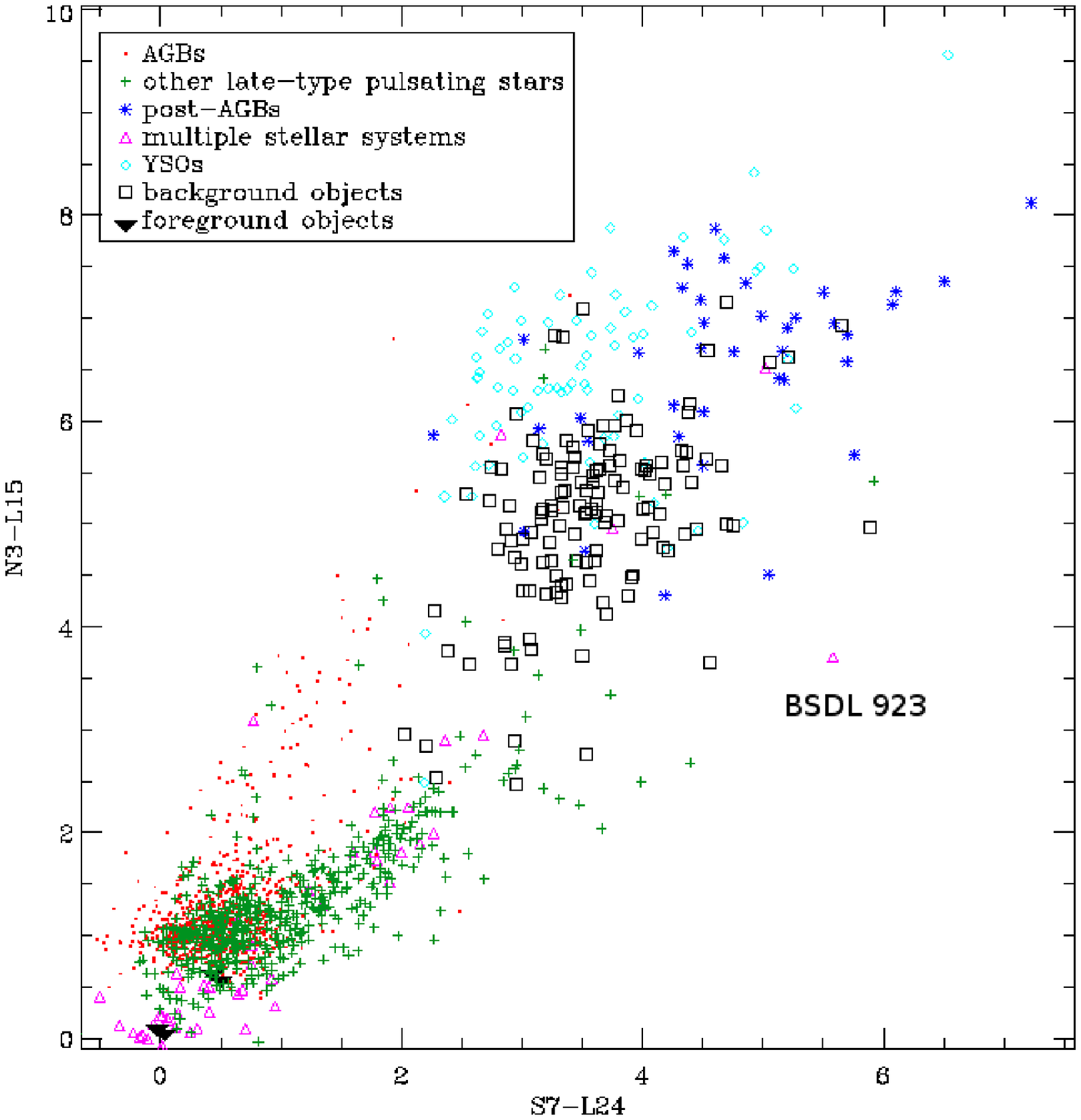}}
\centerline{\includegraphics[width=0.411\textwidth,clip]{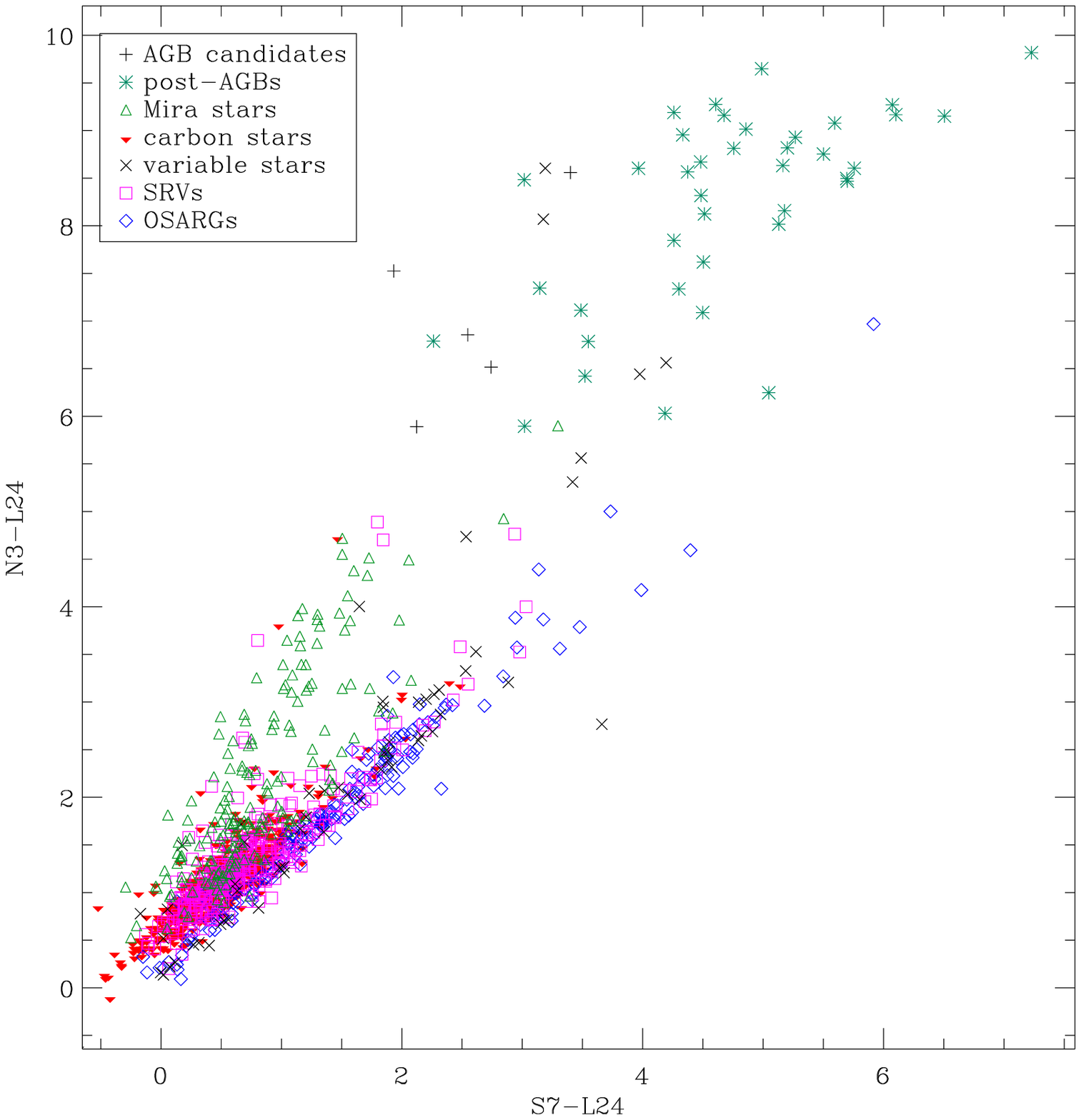}}
\caption{Position of different types of objects in the color-color diagrams $S11-L15$ vs. $N3-S11$(upper), $S7-L24$ vs. $N3-L15$(middle), 
$S7-L24$ vs. $N3-L24$ (lower). 
AGB stars are shown as small gray filled circles, other late-type pulsating giants as gray plus signs, post-AGB stars as asterisks, 
multiple stellar systems as gray open triangles, YSOs as gray diamonds, 
background objects as black open squares, and foreground objects as black filled triangles.\label{fig:cc}}
\end{figure}

As in  Sect.~\ref{r:cmdall}, we divide classified sources into groups 
of: (1) AGB stars, (2) other late-type pulsating giants, (3) post-AGB stars, (4) multiple stellar systems, 
(5) YSOs, (6) background objects, (7) foreground objects. 
Color-color diagrams of the different types of sources from our sample are shown in Fig.~\ref{fig:cc}.

In all color-color diagrams different types are well 
separated. Especially, AGB stars and YSOs tend to occupy 
separate parts of the diagrams. YSOs and post-AGB stars are significantly ``redder'' than most AGB stars or possible AGB stars, which is 
visible on all diagrams. AGB stars with other late-type pulsating giants shows a rather clear pattern on all color-color diagrams with a tendency to occupy the ``bluer'' 
parts of the diagrams. 
As shown in Fig.~\ref{fig:cc}, especially in the middle and lower panels, a significant part of the AGB group forms a clearly separated tail going 
from the ``bluer'' part of the diagram into the ``redder'' part. A small group of post-AGBs tend to be placed in the region of AGB stars, but also seems to remain in its redder part. 
Multiple stellar systems also tend to occupy a separate region, a significant amount form a tight ``cloud'' in the ``bluer'' part 
of the diagram, which is visible in the middle panel of Fig.~\ref{fig:cc}. 
However, a group of multiple systems in nonuniform and single 
objects are spread over the whole diagram. 

One of unusual sources in this group of multiple stellar objects corresponds to \texttt{BSDL~923 (Dachs LMC 1-11, SSID69)} classified in the SIMBAD database as 
a cluster. However, elsewhere, this object is classified as a massive star belonging to a young stellar cluster
LMC-N30 [Gouliermis et al., 2003, Bica et al., 1999], or as a B supergiant; hence, it is classified as OTHER in a new classification of 197 point sources observed 
with the Infrared Spectrograph in the SAGE-Spec Legacy program on the Spitzer Space Telescope [Woods et al., 2011].
For example, in the diagram $S7-L24$ vs. $N3-L15$ (middle panel of Fig.~\ref{fig:cc}) it is the only object classified as a multiple stellar system  
in the ``redder'' part of the diagram. Taking into consideration the dense cluster environment of this source, it is possible that our classification is not correct. 
Thus, unusual objects which are 
placed in a nonstandard place in the color-color diagrams, tend not to be of a clearly defined type.  

As mentioned in Sect.~\ref{r:data}, only 4 objects within the Milky Way Galaxy were found. They occupy a separate region in the ``bluer'' part of the 
color-color diagrams. Because of the small number of objects, it is difficult to define any characteristic tendency.
Also, background objects tend to occupy separate regions in color-color diagrams, which can especially be seen in Fig.~\ref{fig:cc}. 
This group forms a ``cloud'', which is 
``redder'' than the AGB group, but ``bluer'' than YSOs. 
Following Ita et al. [2008], we can distinguish two main groups of different objects in the diagram $S11-L15$ vs $N3-S11$. One is centered around (0.1,1) and the second 
around (1.0, 4.0). Ita et al. [2008] further separate the first group into sources without circumstellar dust (centered around (0.2,0.2)) and red giants with circumstellar 
emission (centered around (0.25,0.8)). Also, it can be visible in the diagram, as suggested by Ita et al. [2008], that red giants become ``bluer'' and turn red again 
during their evolution, which may be caused by changes of 
strength of silicate, or silicon carbide, dust features in the S11 band relative to the L15 band [Ita et al., 2008]. 
Ita et al. [2008] separate a large group of YSO candidates, 
including PPNe and PNe, which satisfies the condition 
($N3-S11$) $>$ 2.0 and ($S11-L15$) $>$ 1.0. Also, background galaxies fulfill this restriction [Ita et al., 2008]. This condition also restricts our group to 
YSOs and background objects. Moreover, we can separate 
a group of YSOs (contaminated by some PNe) in the diagram $S11-L15$ vs $N3-S11$, shown in the upper panel of Fig.~\ref{fig:cc}, by the following conditions:
\begin{itemize}
 \item $N3-S11$ $>$ 4.5,
 \item $S11-L15$ $>$ 0.5.
\end{itemize}

\subsection{Separation of AGB and post-AGB stars in color-color diagrams}\label{r:coloragb}

\begin{figure}[t]
\centerline{\includegraphics[width=0.411\textwidth,clip]{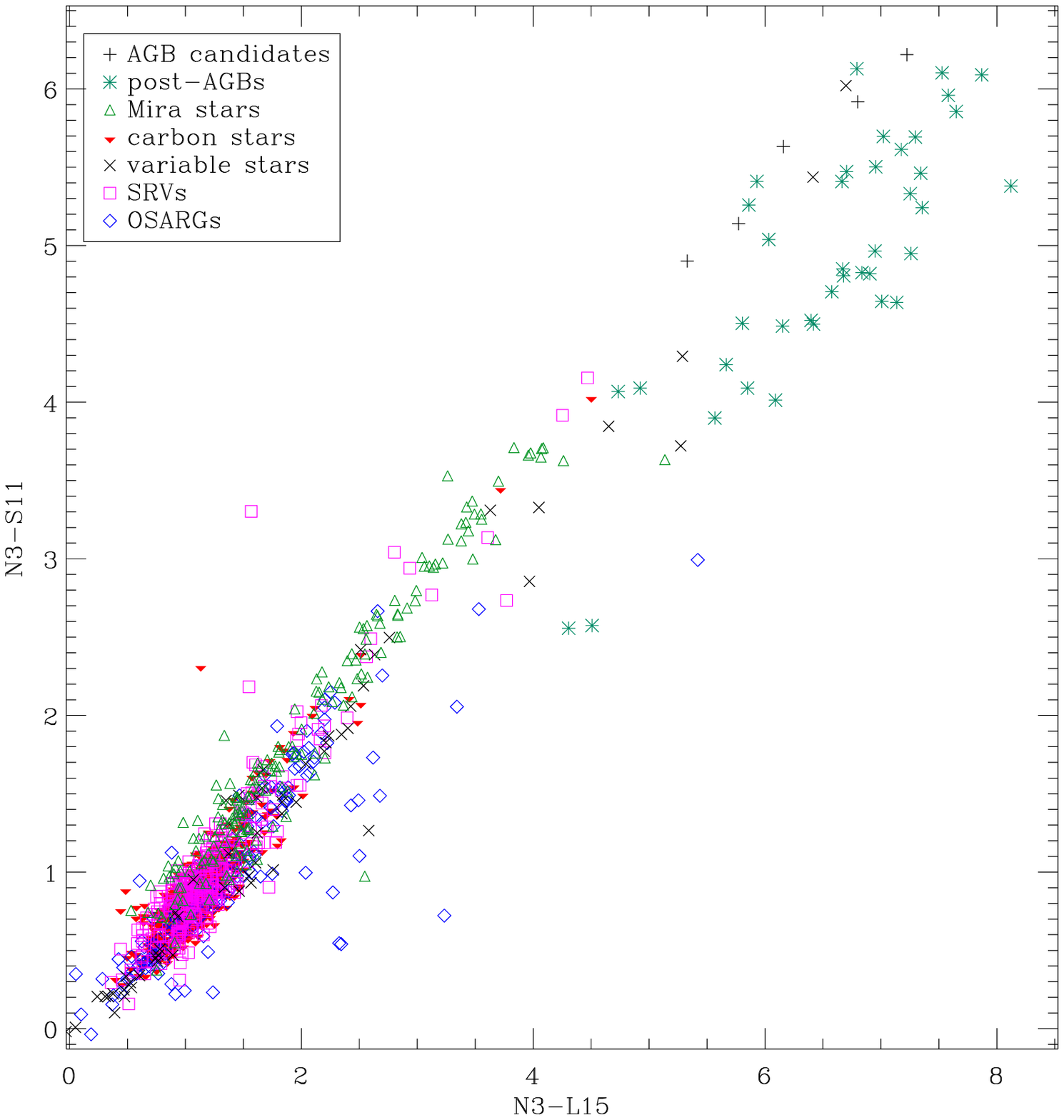}}
\centerline{\includegraphics[width=0.411\textwidth,clip]{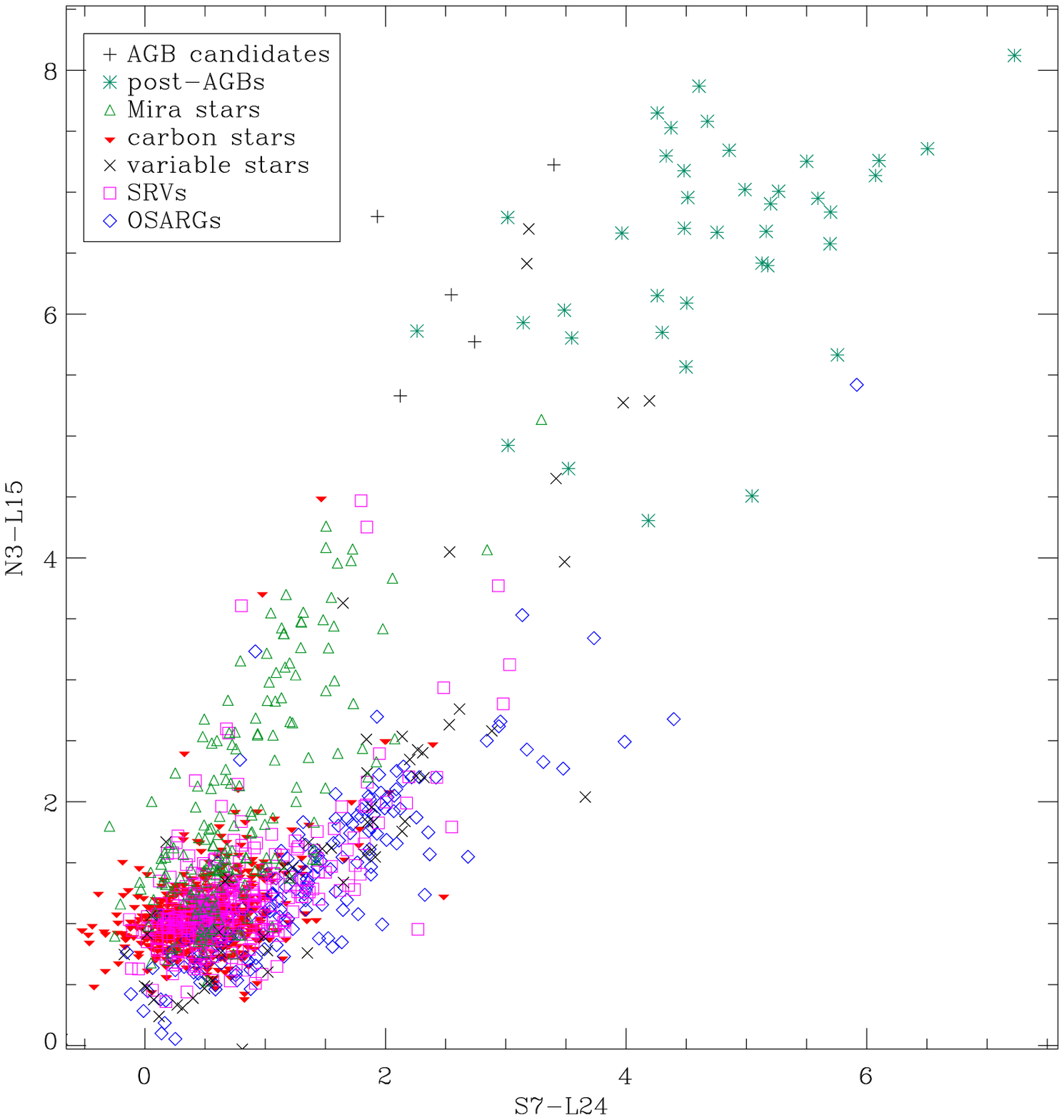}}
\centerline{\includegraphics[width=0.411\textwidth,clip]{N3-L24vsS7-L24.eps}}
\caption{Position of different types of objects in the color-color diagrams $N3-L15$ vs. $N3-S11$ (upper), $S7-L24$ vs. $N3-L15$ 
(middle), $S7-L24$ vs. $N3-L24$ (lower). 
(d) AGB candidates  are shown as plus signs, (a) post-AGB stars as gray asterisks, 
(b) Mira stars as gray open triangles, (c) carbon stars as black filled triangles, (e) variable stars as X signs, 
(f) SRVs as gray open squares, and (g) OSARGs as diamonds.\label{fig:ccagb}}
\end{figure}

Color-color diagrams with near- and mid-infrared wavelengths are a useful tool for discriminating various types of AGB, and post-AGB, stars. 
In the next step, we analyze AGB, and post-AGB, stars from our sample selecting, as in Sect.~\ref{r:cmdagb}, the following types: (a) post-AGB stars (PNs), (b) Mira stars, 
(c) carbon stars, (d) candidates for AGB stars, (e) variable stars, (f) semiregular variables (SRV), (g) OGLE small amplitude red giants 
(OSARG). Color-color diagrams with different types of AGB, and post-AGB, stars are presented in Fig.~\ref{fig:ccagb}. 
In all diagrams, carbon 
stars form a clear cloud in the ``bluer'' parts of the diagram. Also, post-AGBs,  which occupy the ``redder'' part, are well separated. Mira variables create a long tail going 
from carbon stars, and become ``redder''.  Candidates for AGBs form a cloud around carbon stars. 
However, AGB stars appear in a cloud of 
carbon stars while candidates for AGBs tend to form an adjacent group to carbon stars, but a little bit ``bluer''. 

\section{Background objects and foreground Milky Way sources}\label{r:cmdforeback}

We distinguish a group of foreground and background objects in our sample during the analysis. Stars denoted as objects within the Milky Way Galaxy in the NED database were 
classified as foreground objects. We have found only 4 sources fulfilling this criterion, although we expected to find more. As background objects, we selected 
133 objects classified as a galaxy or QSO or AGN. 
\begin{figure}[t]
\centerline{\includegraphics[width=0.411\textwidth,clip]{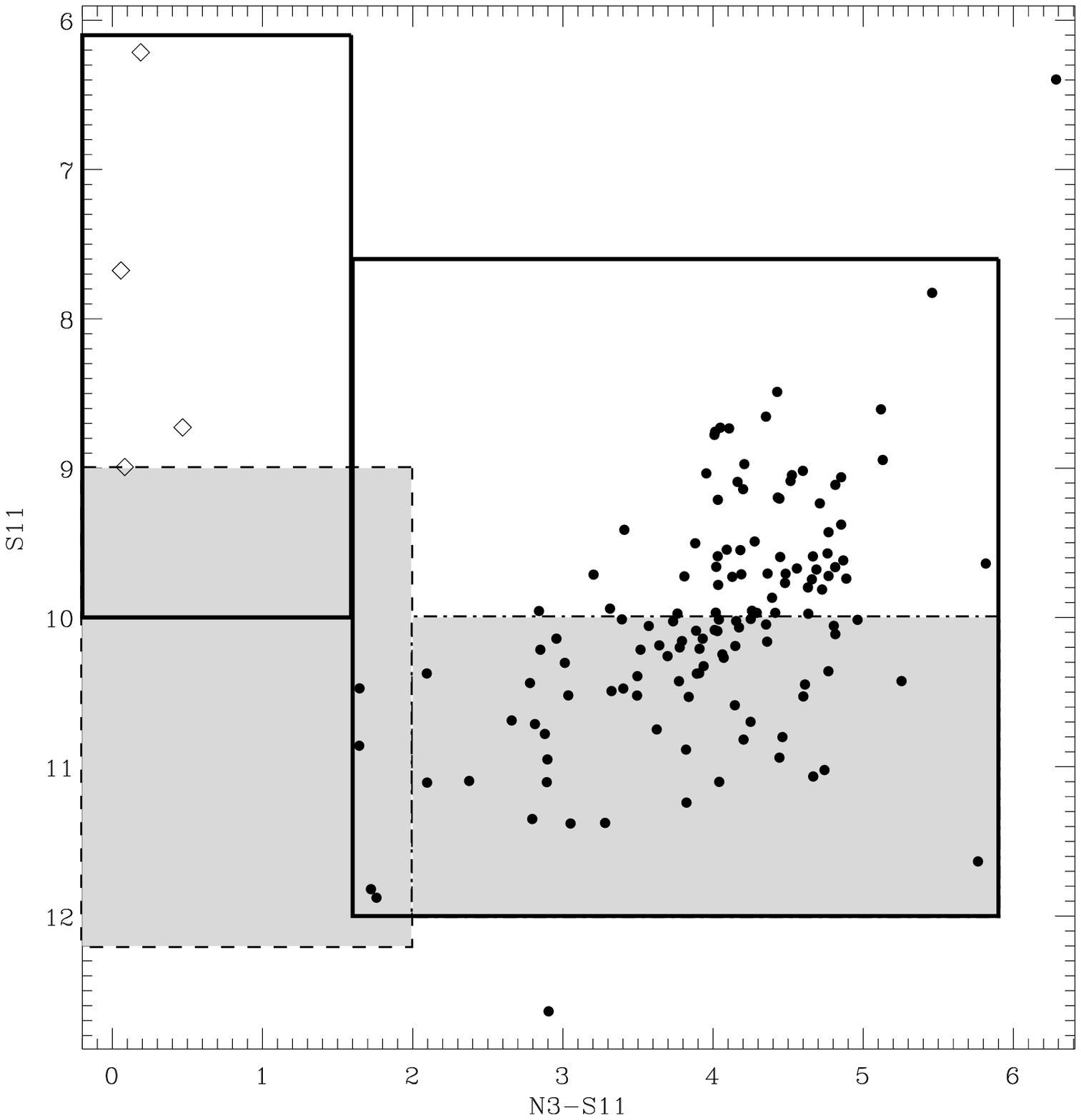}}
\caption{Color-magnitude diagrams $N3-S11$ vs S11 (upper) for sources identified as foreground (diamonds) and background objects (filled circles). 
Objects fulfilling conditions for foreground objects proposed by Ita et al. [2008] are placed in the gray area with a dashed outline, objects corresponding to background 
galaxies have a dash-dotted outline. Modifications of the conditions adapted for our sample is shown with a solid thick outline. \label{fig:back}}
\end{figure}
As shown in the color-magnitude diagrams $N3-S11$ vs S11 with identified foreground and background objects, 
presented in Fig.~\ref{fig:back}, not all sources fulfill the conditions proposed by Ita et al. [2008]. Our sample of background objects includes not only galaxies but also 
AGNs and QSOs, but it is a small group. Taking into consideration the small amount of foreground objects, it is difficult to propose any separation. 
Based on diagram $N3-S11$ 
vs. S11 for our sample, following modifications of these conditions, the following can be proposed:
\begin{enumerate}
 \item Background objects:
\begin{itemize}
 \item $N3-S11 >$ 1.6 [mag],
 \item 7.6 $<$ S11 [mag] $<$ 12.
\end{itemize}
 \item Foreground objects;
\begin{itemize}
 \item $N3-S11 <$ 1.6 [mag],
 \item 6.1 $<$ S11 [mag] $<$ 10. 
\end{itemize}
\end{enumerate}

However, Ita et al. [2008] expected to detect about 500 background galaxies and 190 foreground objects in a square degree and the fraction of foreground stars was estimated to 
be about 13$\%$. In our sample, less than 0.1$\%$ are foreground objects. Thus, we searched for 
candidates for Galactic objects taking into consideration sources with z $<$ 0.000927\footnote{$^{3}$Information about the values of z is taken from the SIMBAD database}. This condition was met by 132 objects, mostly classified as other (60 $\%$). 
Among these, 47 sources satisfied the conditions proposed by Ita et al. [2008], and this bacame 113 objects after taking into consideration our modifications. 
\begin{figure}[t]
\centerline{\includegraphics[width=0.411\textwidth,clip]{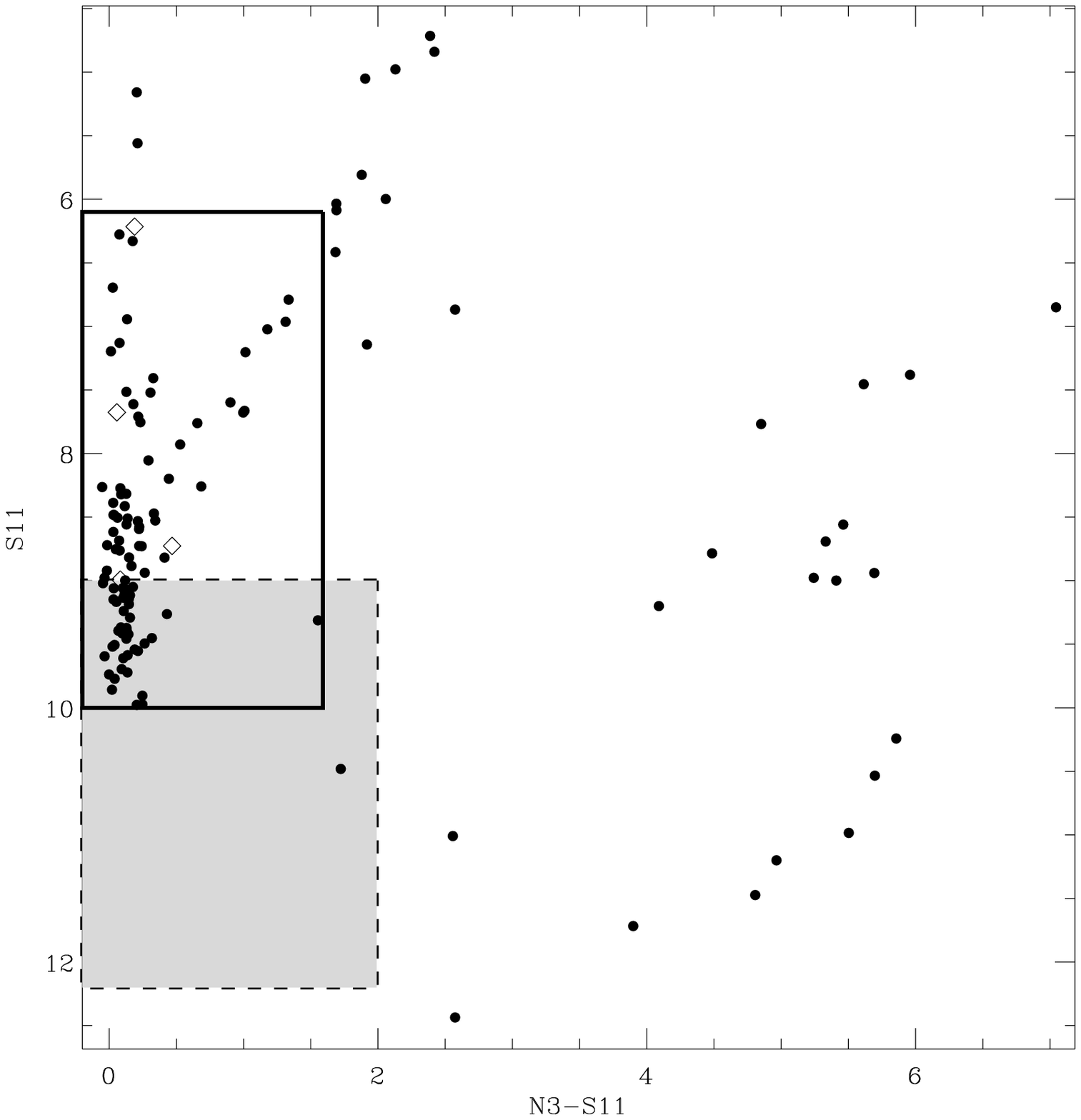}}
\caption{Color-magnitude diagrams $N3-S11$ vs S11 for sources suspected of being within Milky Way (filled circles) and classified as 
foreground objects (diamonds). 
Objects fulfilling conditions for foreground objects proposed by Ita et al. [2008] are placed in gray area with dashed outline, are corresponding for background 
galaxies have dash-doted outline. Modifications of conditions adapted for our sample is shown with solid thick outline. \label{fig:z}}
\end{figure}

\section{Sources without counterparts}\label{r:cmdnotident}
For our analysis, we have selected 3 852 sources with complete five-band color information. However, for almost a quarter of them, no counterparts in the 
NED, SIMBAD, OGLE, 2MASS or AKARI-FIS databases were found (1 016). Moreover, for 872 sources with identified counterparts, a type of object was not determined. In this 
section, we present further analysis of these sources (1 888) and we discuss the color-magnitude and color-color diagrams. In order to determine the possible types and properties 
of these sources, we estimate standard fields of different objects at infrared diagrams and discuss the alignment of sources not identified. 

To assign different types of sources to 
different areas of diagrams, we used the classification algorithm from Scikit-learn [Pedregosa et al., 2011] based on a Support Vector Machine (SVM) [Chang and Lin, 2011]. 
This is a useful tool used for classification and 
regression purposes, e.g., to classify structures in the Interstellar Medium [Beaumont et al., 2011], or AGNs from stars and normal galaxies [Zhang et al., 2002], 
or to select QSO candidates from the MACHO LMC database [Kim et al., 2011a; Kim et al., 2011b]. 

For our data, a binary classification was performed 
using a non-linear SVC with a radial basis function (RBF) kernel. The SVM was trained on 
identified sources. The obtained separation conditions of different group of objects were then used to classify the unknown sources.

\subsection{Candidates for foreground and background objects within unidentified sources}\label{r:cmdnotidentfore}

Within a sample of unidentified sources, we try to find candidates for background objects and sources within the Milky Way. Our analysis is based on the properties of the 
diagram N3-S11 vs S11. As a training group to determine separate reference fields, we used sources identified as background, or Galactic, objects. Objects suspected 
of being foreground or background objects, presented in Sect.~\ref{r:cmdforeback}, are not taken into consideration. 

\begin{figure}[t]
\centerline{\includegraphics[width=0.411\textwidth,clip]{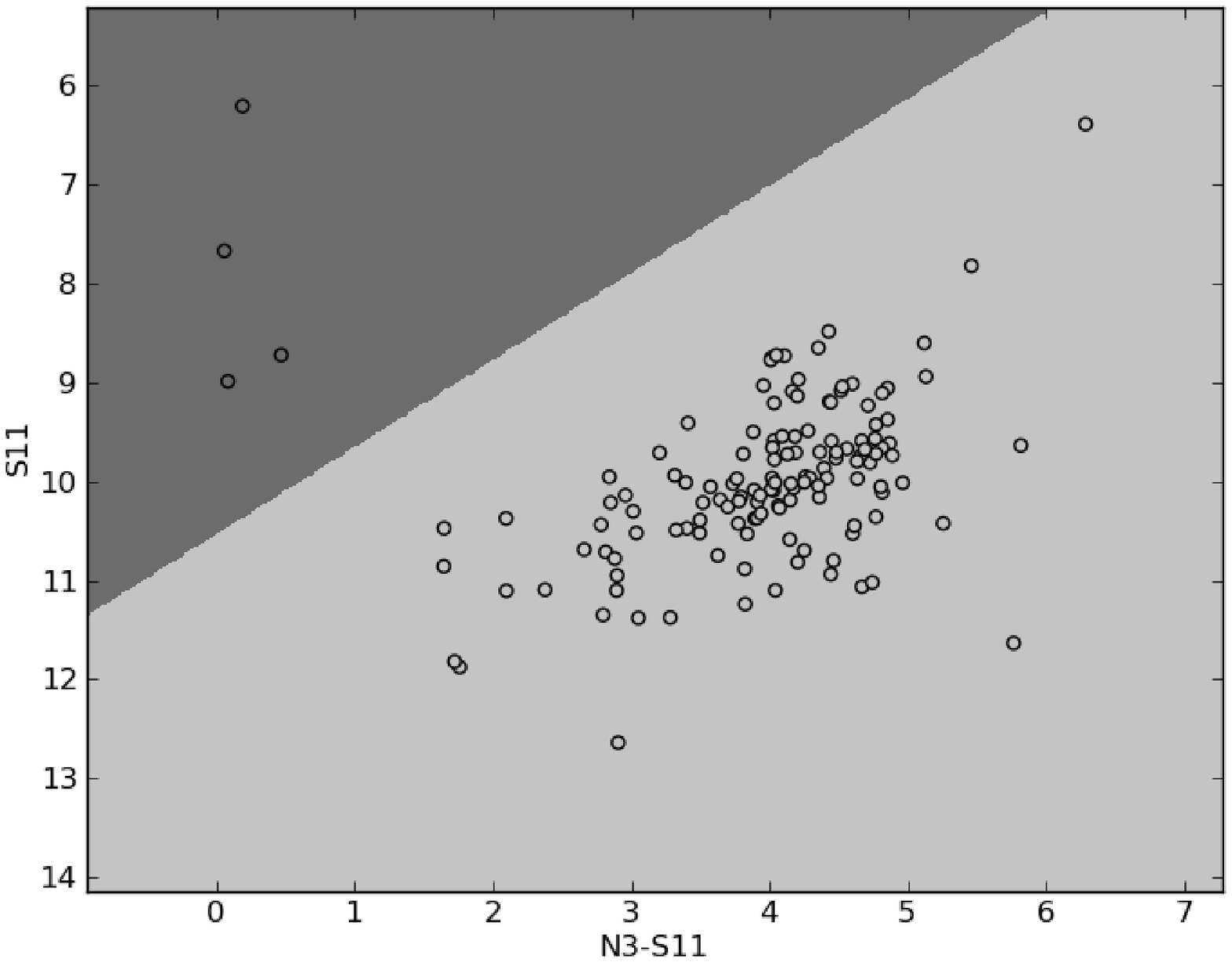}}
\centerline{\includegraphics[width=0.411\textwidth,clip]{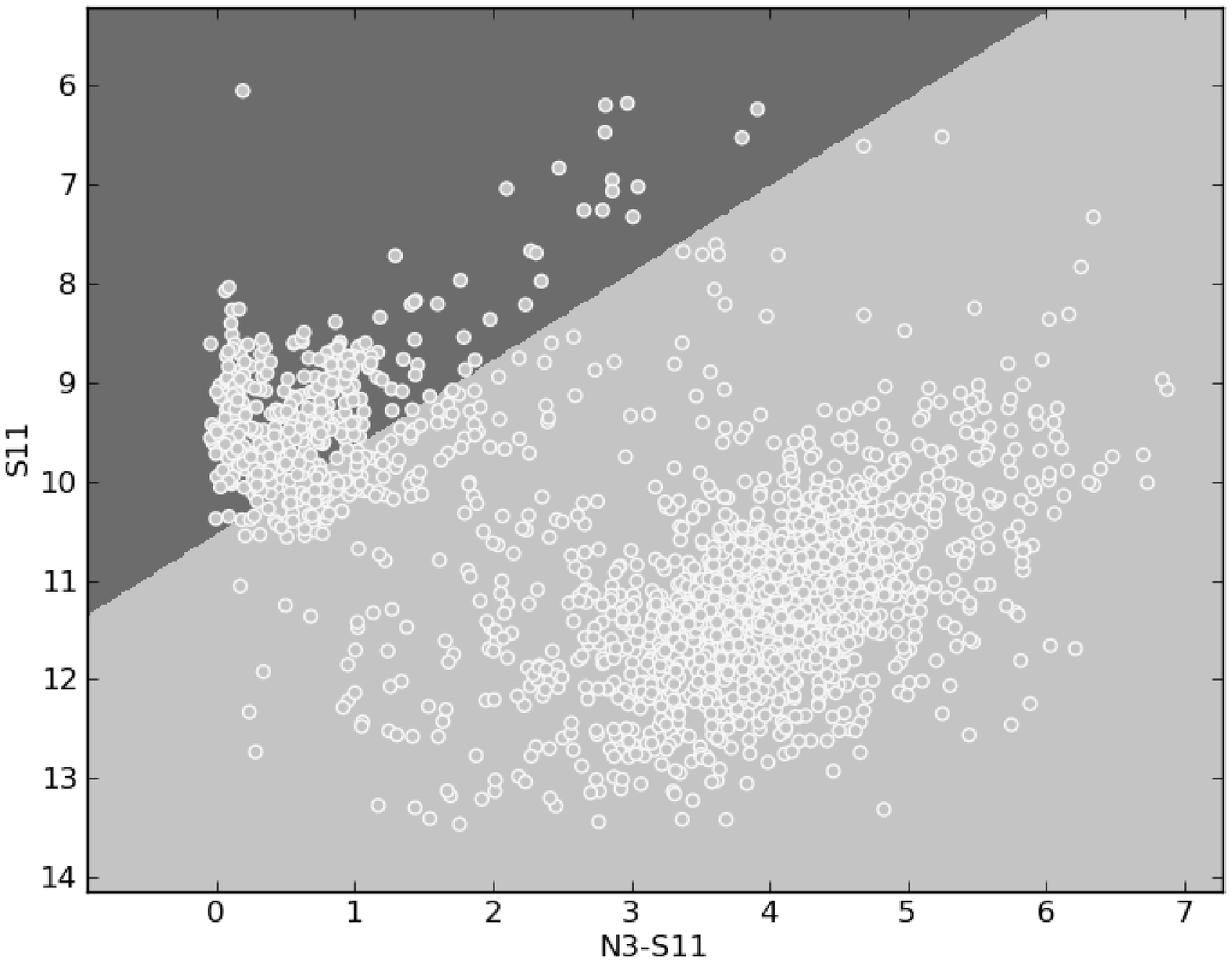}}
\caption{Color-magnitude diagram $N3-S11$ vs S11 for sources identified as background, or foreground, objects (upper), and not identified (lower) with the classification 
based on SVM. The dark gray area corresponds to sources within the Milky Way, and the light gray area to background objects.  \label{fig:not_identified_fore}}
\end{figure}

The color-magnitude diagram $N3-S11$ vs S11 is presented in Fig.~\ref{fig:not_identified_fore}. In the upper panel, data used for training are plotted, 
and in the lower panel 
unidentified sources are plotted. There is a small amount of identified (upper panel) foreground objects (dark gray area) and a large amount of background objects 
(light gray area). We found a similar tendency in this case of the unidentified sources. A dominant number of sources without counterparts are placed in the light gray area,  
corresponding to background objects.  
 
\subsection{Separation of sources without counterparts in color-magnitude diagrams}\label{r:cmdnotidentcc}

\begin{figure}[t]
\centerline{\includegraphics[width=0.411\textwidth,clip]{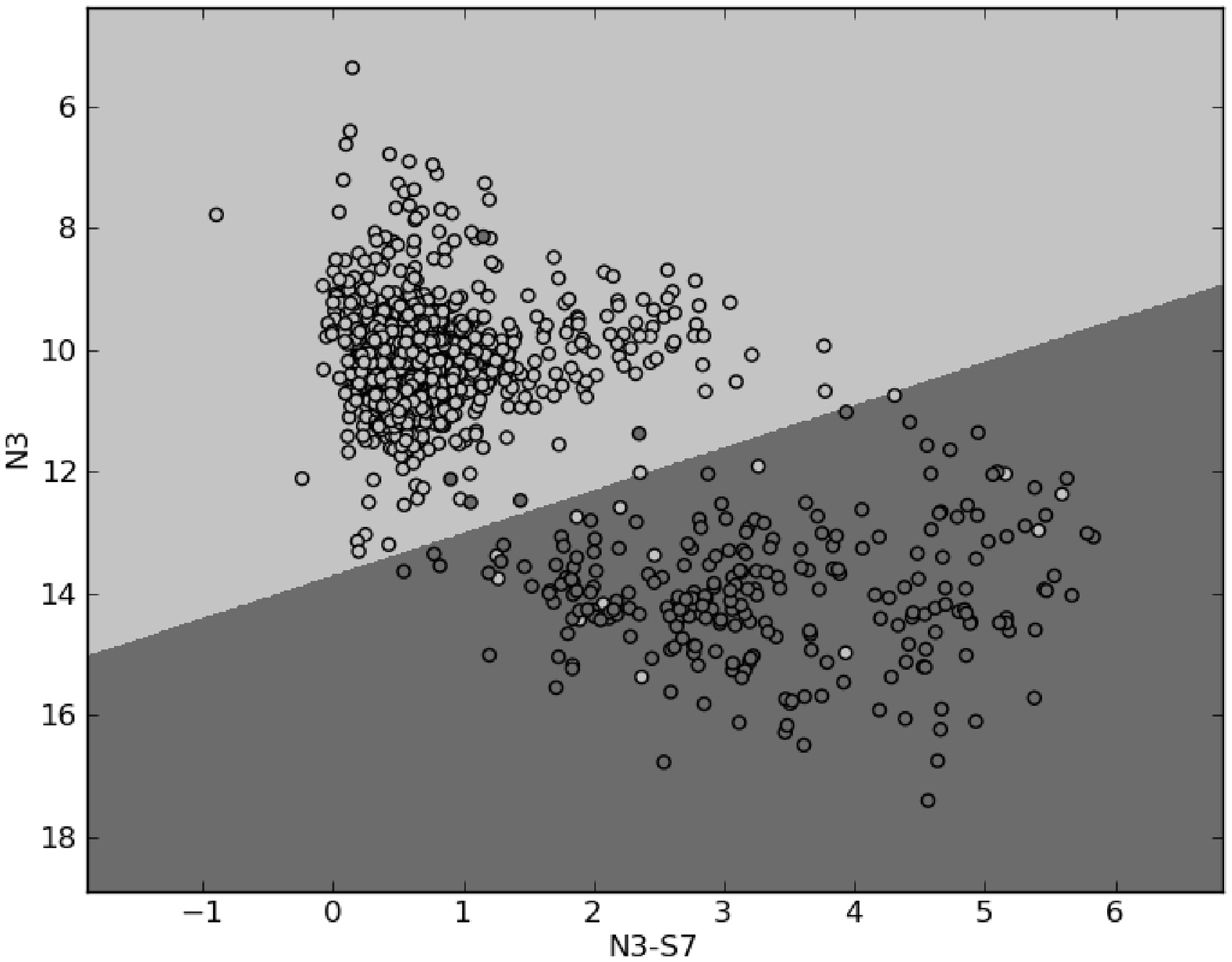}}
\centerline{\includegraphics[width=0.411\textwidth,clip]{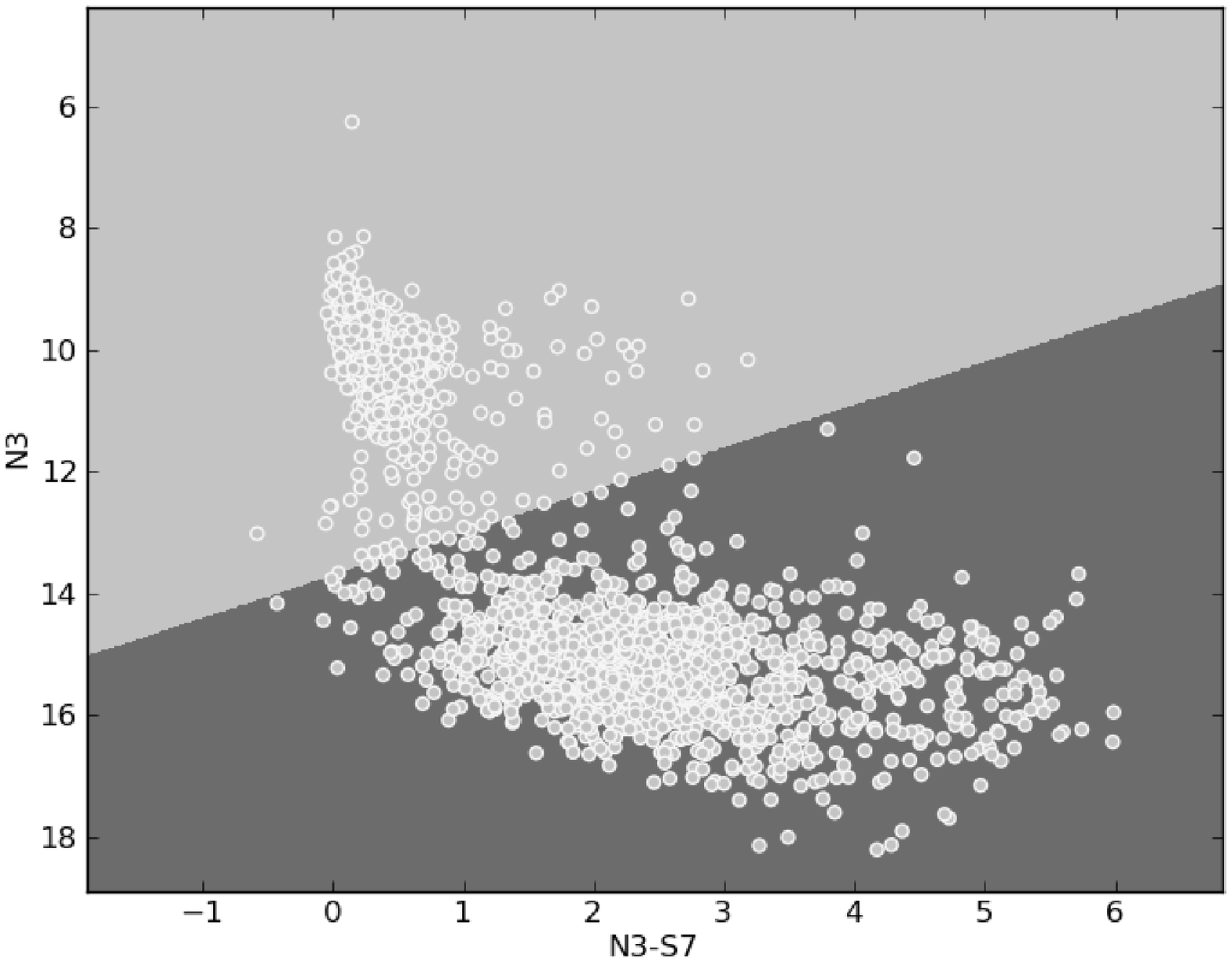}}
\caption{Color-magnitude diagram $N3-S7$ vs N3 for sources identified as (1) foreground objects, AGBs, possible AGB stars, and multiple stars (light gray area) or (2) 
post-AGBs, YSOs and background objects (dark gray area)(upper panel) and not identified (lower panel) with classification 
based on SVM. The dark gray area corresponds to objects classified as foreground objects, AGBs, possible AGB stars, or multiple stars. The light gray area corresponds to 
post-AGBs, YSOs, and background objects.  \label{fig:not_identified_cmd}}
\end{figure}

As shown in Sect.~\ref{r:cmd}, based on color-magnitude diagrams, we can see separation between different types of objects. We tried to classify not identified sources 
using a diagram $N3-S7$ vs S7.  In Fig.~\ref{fig:cmd} (upper panel), we can see a clear separation between two clouds: one formed by foreground objects, AGBs, and possible 
AGB stars, and multiple stellar systems; and a second one consisting of post-AGBs, YSOs, and background sources. We used these two groups to train the SVM and the result is 
presented in the upper panel of Fig.~\ref{fig:not_identified_cmd}.
Within a sample the ``bluer'', and more luminous, group is significantly larger than the ``redder'' one (1 409: 243). Taking into consideration sources
not identified, this 
dependence is opposite, i.e., more sources tend to be post-AGBs, YSOs, or background objects.    

\section{Contribution from main sources to the total infrared flux in LMC}\label{r:flux}

Observations made using the Spitzer Space Telescope have enabled gas and dust input from AGB stars to be studied in the LMC [Matsuura et al., 2009]. 
The dominant source of gas are AGBs 
and possibly supernovea; however, present-day star formation depends on gas already present in the ISM. Thus, the star-formation rate will probably decline, 
when gas in the ISM 
becomes exhausted, unless another gas feedback is provided [Matsuura et al., 2009]. Matsuura et al. [2009] suggest that the LMC have common for high-z galaxies `a missing 
dust-mass problem'. Dust mass contributed from AGBs and SNe over the dust life time is significantly less than the dust mass in the ISM, so another source of dust is required 
[Matsuura et al., 2009]. AGBs, especially those carbon-rich, are a dominant source of gas and dust for the ISM of the LMC. 
Dust production from SNe is very uncertain; however, 
SNe are important contributors of gas. Also, novae and R CrB stars can expel dust and gas to the ISM, however their feedback is minor [Matsuura et al., 2009]. 

\begin{figure}[t]
\centerline{\includegraphics[width=0.5\textwidth,clip]{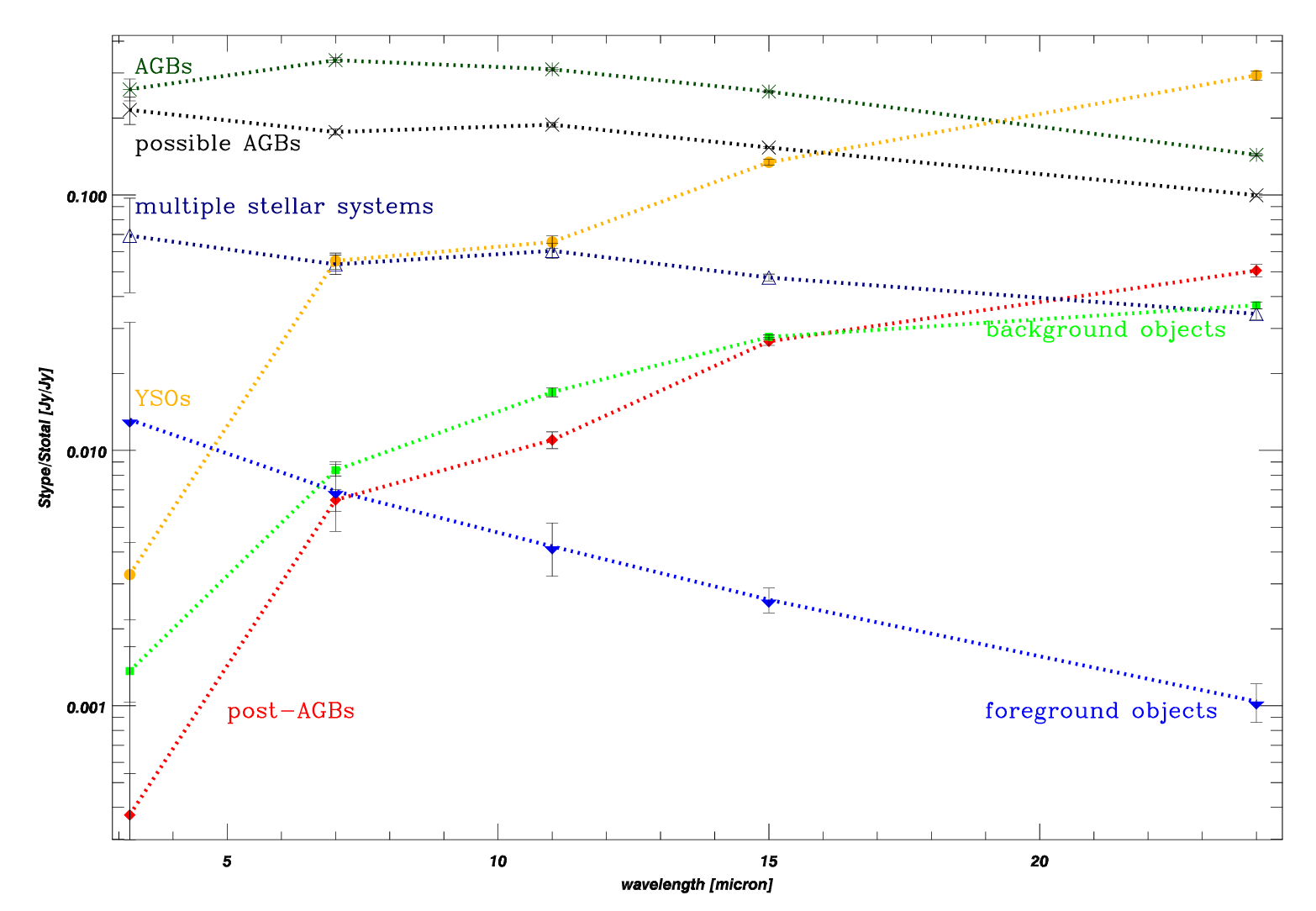}}
\caption{Contribution of different types of objects to the total number of point sources detected in NIR and MIR at each band ($S_{total}$). AGB stars are shown as asterisks, late-type pulsating 
giants as X sign, multiple stellar systems as triangles, YSOs as diamonds, post-AGB as filled diamonds, background objects as squares, foreground objects 
as filled triangles, other as circles, sources of an unknown origin as filled squares, and sources not identified as filled circles.
\label{fig:luminosity}}
\end{figure}

The contribution from different objects to the total NIR and MIR flux is shown in Fig.~\ref{fig:luminosity}. 
In all the NIR and MIR wavelengths, AGB stars are the dominant source of LMC radiation from point sources, which corresponds to their number percent. 
YSOs, and post-AGBs, are more visible at longer wavelengths, while AGBs contribute more and more when we move to shorter wavelengths. Objects with an unknown origin also, as 
post-AGBs and YSOs, contribute more at longer  wavelengths. 
The luminosity contribution of foreground objects falls mainly within the wavelength range 3 to 24 $\mu$m, contrary to the contribution from background objects.

\section{Summary}\label{r:summary}

The AKARI IRC survey performed observations over an area of about 10 $\rm deg^2$ 
of the Large Magellanic Cloud. We prepared a catalog of counterparts based on the Release Candidate 
version 1 of the point source catalog of the AKARI LMC Large Area Survey. 
Our analysis is based on sources detected at all NIR  and MIR wavelengths of 3.2, 7.0, 11.0, 
15.0, 24.0 $\mu$m, i.e. with a complete five-band color information. 
We cross-correlated these sources with publicly-available databases at different wavelengths and found 
counterparts for almost 75~$\%$ of our sample. A majority (nearly 70~$\%$)
of identified sources are AGB stars and similar evolved giants. 
Consequently, AGB stars are the dominant source of the LMC radiation at 
all NIR and MIR wavelengths from point sources. Their luminosity contribution rises with wavelength 
from 3 to 24 $\mu$m, contrary to the contribution from YSOs and PNe. 
This IR budget should be useful for future modeling of detailed SEDs of LMC-like  irregular galaxies.
In this paper, we have also presented color-magnitude, and color-color, diagrams showing 
new interesting features. We have distinguished different groups of AGBs, and post-AGBs,  
and discussed color-magnitude, and color-color, diagrams. Within a sample there is 
a large number of sources without counterparts or of an unidentified type, for which we present 
color-magnitude diagrams. Using an SVM, we have proposed a method to determine the properties 
of unidentified objects from their infrared color information.

\acknowledgments{

We thank both referees for providing constructive comments
and help in improving this paper.
This work is based on observations with AKARI, a JAXA project with the participation of ESA. 
This research has made use of the NASA/IPAC Extragalactic Database (NED) which is 
operated by the Jet Propulsion Laboratory, California Institute of Technology, under contract with
the National Aeronautics and Space Administration, and the SIMBAD database, operated at 
CDS, Strasbourg, France. 

AP and MS have been supported by the research grant of the Polish
Ministry of Science N N203 51 29 38. This research was partially supported by 
the project POLISH-SWISS ASTRO PROJECT co-financed by a grant from Switzerland 
through the Swiss Contribution to the enlarged European Union. 

TTT has been supported by the 
Grant-in-Aid for the Scientific Research Fund (20740105, 23340046, and
24111707) and for the Global COE Program Request for Fundamental
Principles in the Universe: from Particles to the Solar System and the 
Cosmos commissioned by the Ministry of Education, Culture, Sports, 
Science and Technology (MEXT) of Japan.}

\email{M. Siudek (e-mail: gsiudek@cft.edu.pl)}
\label{finalpage}
\lastpagesettings

 \include{appendix}

\lastpagecontrol{20cm}

\end{document}